\documentclass{aastex}
\pdfoutput=1

\begin{document}


\title{Matched photometric catalogs of GALEX UV sources with Gaia DR2 and SDSS DR14 databases (GUVmatch)} 


\author{Luciana Bianchi\altaffilmark{1} and  
  Bernard Shiao \altaffilmark{2}}
\altaffiltext{1}{Dept. of Physics \& Astronomy, The Johns Hopkins University, 3400 N. Charles St.,  Baltimore, MD 21218, USA; 
http://dolomiti.pha.jhu.edu}
\altaffiltext{2}{Space Telescope Science Institute, 3400 San Martin Dr., Baltimore, MD 21210} 
\email{bianchi@jhu.edu}

\begin{abstract}
We have matched the ultraviolet (UV) sources in $GUVcat$\_AIS \citep{bia17guvcat,bia20guvcat}  with optical databases having similar depth and wide sky coverage.  $GUVcat$\_AIS has GALEX far-UV (FUV, $\lambda$$_{eff}$ $\sim$ 1528\AA )  and near-UV (NUV, $\lambda$$_{eff}$ $\sim$2310\AA)  photometry of $\approx$83~million 
	sources, covering 24,788~square~degrees of the sky, with typical depth of FUV=19.9, NUV=20.8~ABmag.   Matches with Gaia and SDSS databases are presented here.

 Gaia data~release~2 (DR2),  covering the entire $GUVcat$ footprint \citep{bia19areacat},
 detected about one third of the $GUVcat\_AIS$ sources. We found 31,925,294   Gaia~DR2 counterparts
to 30,024,791  $GUVcat$\_AIS unique sources, with photometry in $Gaia$~$G$ band, and often also in $Gaia$~$BP$ and $RP$ bands; 
26,275,572  matches have a parallax measurement,  21,084,628~/~18,588,140~/~16,357,505 
with parallax error less than  50\%/30\%/20\%. 

The match with SDSS data release 14 (DR14)  yields 23,310,532  counterparts to 22,207,563  unique $GUVcat$\_AIS sources, 10,167,460 
of which are point-like, over a total overlap area of $\approx$11,100~square~degrees \citep{bia19areacat}. SDSS adds to the UV photometry five optical magnitudes: $u,~g,~r,~i,~z$, and optical spectra of 860,224 matched sources. 

We used  a match radius of 3{\mbox{$^{\prime\prime}$~}}, consistent with previous works (e.g., \citet{bia11a}), although the positions agree to $\lesssim$1.5{\mbox{$^{\prime\prime}$~}}  for the majority of [point-like] matched-sources, in order to identify possible multiple matches whose UV flux could be unresolved in GALEX imaging. The catalogs can be trimmed to a tighter match radius using the provided separation. 

 The multi-band photometry is used to identify classes of astrophysical objects that are prominent in UV, to characterize the  content of the $GUVmatch$ catalogs, where stars in different evolutionary stages, QSOs, and galaxies can be separated.

\end{abstract}


\section{Introduction. }
\label{s_intro}

\subsection{The catalog of UV sources  $GUVcat$}
\label{s_introguvcat}

To facilitate statistical studies of UV sources, \citet{bia17guvcat}(revised 2020)
constructed $GUVcat\_AIS$, a catalog of $\sim$83~million GALEX\footnote{The Galaxy Evolution Explorer (GALEX, \citet{martin05}) operated from July 2003 to Feb. 2012, surveying the sky simoultaneously in FUV and NUV, until the FUV detector stopped working in 2009. A total of almost 600~million individual source measurements are in the GALEX database  \citep{bia09,bia11,bia14}} unique UV sources, by removing duplicate measurements from repeated observations of the same source.  
The GALEX All-sky Imaging Survey (AIS)  consists of 57,000 distinct imaging observations (``visits'') of over 28,700 fields. To build 
 $GUVcat$\_AIS, \citet{bia17guvcat} used both ``co-adds'' and ``visits'', in order to cure some incorrect associations (``co-adds'')  of same-field visits in the original GALEX pipeline processing.  See \citet{bia17guvcat} for details, and for description of other caveats. $GUVcat$\_AIS supersedes earlier versions of unique UV source catalogs constructed from previous  data releases, such as the widely used $BCScat$ \citep{bia14uvsky} and the earlier catalogs of   \citet{bia11a, bia11b}. 

 In $GUVcat$  only one measurement for each UV source is retained (unique source list), and  the field edges, that are plagued with artifacts, are trimmed. $GUVcat$\_AIS\_FOV055 contains all sources within 0.55{\mbox{$^{\circ}$}}~ of the field center, and eliminates most rim artifacts without loss of real sources (cfr \citet{bia17guvcat}). 
 $GUVcat$, alone or matched with surveys at other wavelengths,  enables effective selection of UV-source samples (e.g., \citet{bia07apj,bia11a,bia18b}). 
The power to identify and characterize astrophysical sources is  maximized when the UV measurements are complemented with optical-IR  data 
(e.g., \citet{bia11a,bia11b,bia09,bia09qso,bia07apj,bia05apjl}; \citet{Hutchings2010,Hutchings2010b}; \citet{myron}). 
In order to match two databases, unique-source catalogs must be used for each (no repeated entries for the same source). Currently, for GALEX  this is only possible  using $GUVcat$,
while the original database\footnote{hosted at MAST: \url{https://galex.stsci.edu} } includes repeated measurements, that are instead useful for  serendipitous variability searches (e.g., \citet{conti14,alexbia} and references therein). 
 $GUVcat$\_AIS includes all fields observed with both FUV and NUV detectors in the All-sky Imaging Survey (AIS), the survey with the widest sky coverage (see  Figure 1 of  
\citet{bia17guvcat}, and \citet{bia14uvsky} for more maps; also \citet{bia09} for an overview). By excluding the observations that have only the NUV detector exposed (cfr Figure 1-left of \citet{bia17guvcat}) 
we ensure that the statistics of samples selected by criteria including the FUV-NUV color  is unbiased.
 
   $GUVcat$\_AIS\_FOV055 (\citet{bia17guvcat},  revised by \citet{bia20guvcat}) contains 82,992,086 unique UV sources, over an area of 24,790~square degrees \citep{bia19areacat}. Coverage is scant at low Galactic latitudes, due to the abundance of UV-bright hot stars near the Galactic plane that would violate the countrate safety limits of GALEX detectors, as shown in Figure 1 of \citet{bia17guvcat}. 
The tags (columns)  included in $GUVcat$\_AIS are described in Table 8 of \citet{bia17guvcat}. Among the tags constructed in addition to those provided by the GALEX pipeline, we recall two of interest here: INLARGEOBJ and LARGEOBJSIZE. They indicate whether a UV source is inside the footprint of an extended object such as a nearby galaxy or dense stellar cluster\footnote{if so, INLARGEOBJ is not 'N' and LARGEOBJSIZE is $>$0.}  (and, if so,  the size of the extended object), where quality of source detection, characterization and measurement can be highly compromised, as illustrated clearly in Figure 5 of \citet{bia17guvcat}.

  $GUVcat\_AIS$ has a typical depth of FUV=19.9, NUV=20.8~ABmags. The fraction of NUV sources that have a significant detection also in FUV varies between a few  \% and $\sim$15\% overall from low to high Galactic latitudes, but it also varies significantly  with source brightness 
\citep{bia11b,bia17guvcat}. 
In total, about 10\% of the 
$\sim$83~million  $GUVcat$\_AIS sources have a significant FUV detection.

 The ratio of FUV detections over NUV detections decreases  towards low latitudes, but it actually increases towards the Galactic plane for the brighter sources, as expected since these are mostly hot stars 
 (see Figure 2 of \citet{bia11b} and Figure 6 of \citet{bia17guvcat}).  The overall decrease is due to the numerous faint sources, mostly extra-Galactic (QSOs for the point-like sources), whose intrinsic density does not depend on Galactic coordinates but whose detection is affected by the severe extinction by dust near the Galactic plane (Figures 2 and 5 of \citet{bia11b}), and whose FUV-NUV color varies with red-shift and therefore also  with observed magnitude (e.g., Section \ref{s_discussion}, also Figures 3-5 of \citet{bia09}, Figure 5 of \citet{bia11a}). 
 UV source densities, by magnitude and by color, at varying Galactic latitudes, are shown for $GUVcat$\_AIS in Figure 6 of \citet{bia17guvcat}, and in form of illustrative sky maps by \citet{bia14uvsky}.

\subsection{Choice of match radius. Accounting for multiple matches.}
\label{s_intromm}

 We matched $GUVcat\_AIS$\_FOV055 with the major optical surveys currently available, the Sloan Digital Sky Survey (SDSS, \citet{sdss}) data release 14  (DR14, \citet{sdssdr14}),  and Gaia data release 2 (DR2, \citet{gaiacoll18}). A forthcoming work will present the match with Pan-STARRS PS1 3$\pi$ survey (\citet{chambersPS16}). For each survey, the characteristics  that are relevant to our purpose are summarized in separate sections below.  

 We note here a general consideration of consequence in any database matching, i.e. the need to  assess the appropriate match radius (maximum distance between source positions in the two catalogs for a source to be retained as a match) such to minimize the number of spurious matches 
 and at the same time minimize the loss of real matches.  These two criteria work in the opposite directions of favouring a more restricted match radius and a more generous one, respectively. Spatial resolution and position accuracy of the databases to be matched determine the optimum match radius, and the density of sources in the sky  affects the incidence of spurious matches.  GALEX's spatial resolution is nominally 4.2/5.2{\mbox{$^{\prime\prime}$~}} (FUV/NUV, \citet{morrissey07});  it degrades towards the field edges. Nearly all sources in the outermost rim, where measurements of position and flux become problematic, are excluded from  $GUVcat$.  The coordinates of the source centroid in GALEX data releases GR6~/~GR7  have a reported accuracy of 0.35/0.48{\mbox{$^{\prime\prime}$~}} (NUV/FUV) in GR6 and  0.32/0.34{\mbox{$^{\prime\prime}$~}} (NUV/FUV) in GR7\footnote{http://www.galex.caltech.edu/researcher/gr6\_docs/GI\_Doc\_Ops7.pdf}.

 Ground-based surveys such as SDSS and Pan-STARRS have a typical spatial resolution of 1.4{\mbox{$^{\prime\prime}$~}}, that is, about 3 times higher than GALEX's, and Gaia has a higher astrometric precision.  On the other hand, UV-emitting sources are very rare, compared with the density of red or IR sources (for both Galactic and extra-Galactic objects), therefore most matches are   unique and reliable (one-to-one).  Nonetheless, due to the higher spatial resolution of the optical databases used here with respect to GALEX, 
 when more than one optical counterpart to a UV source is found within the match radius,  the UV flux could be composite of these optical counterparts, that are resolved e.g. by SDSS or Gaia but unresolved in GALEX imaging. In such cases, keeping only the closest match would be misleading, since the UV flux may be contaminated by a neighbor while the optical flux would be only the flux from the closest match, making UV$-$optical colors meaningless, and the analysis biased.  In the case of multiple matches one can also  look at the magnitudes and colors of the two (or more) counterparts, and if one counterpart is - for example -  much redder and much fainter than another, it is likely that all or most of the UV flux originates from the hot, brighter one. However,  if the  UV source is an unresolved binary comprising a very hot star and a cooler one, the actual optical counterpart may have red optical colors because the flux at longer wavelengths is emitted mostly by the cool companion star (e.g., \citet{bia18b}). 

 To keep track of multiple matches, in order to enable a correct, unbiased scientific exploitation of the matched catalogs, we used a slightly generous match radius of 3{\mbox{$^{\prime\prime}$~}}, which is larger than the typical coordinate offsets between the matched databases, but allows us to identify the cases where two or more nearby optical sources may be unresolved by GALEX.    To facilitate  analysis, we created tags (similarly to \citet{bia11a,bia11b}) by which multiple matches can be immediately identified. The definition of these tags is given in Table \ref{t_mmtags}.

\begin{deluxetable}{ll}
\tabletypesize{\tiny}
\tablecaption{Tags to identify multiple matches within the match radius  \label{t_mmtags} }
\tablehead{
\colhead{Tag} & \multicolumn{1}{c}{Meaning and Values} 
}
\tablewidth{0pt} 
\startdata
DSTARCSEC  &  distance in arcsec between the UV source position and the matched-source position\\
DISTANCERANK  & = 0  : this is the only match to this UV source \\ 
              & = 1  : this is the closest of more than one match to this UV source\\ 
              & =n$>$1 : the n$^{th}$ match to the same UV source, ranked by distance \\
MULTIPLEMATCHCOUNT & \# of matches found for this UV source \\   
                   & (by definition, equal to max(DISTANCERANK) + 1 \\
REVERSEDISTANCERANK\tablenotemark{a} & = 0  if the matched source matches only this one UV source \\ 
                    & = 1  if the   matched source matches also other UV sources, and this is the closest match \\ 
                    &  =n$>$1 : the n$^{th}$ match of this optical source to different UV sources \\
 REVERSEMULTIPLEMATCHCOUNT\tablenotemark{a} & how many UV sources are matched by this optical source  \\   
\enddata
\tablenotetext{a}{these are very rare occurrences, see statistics in Tables \ref{t_statgaia} and \ref{t_statsdss} }
\end{deluxetable}

\begin{deluxetable}{lr}
\tabletypesize{\tiny}
\tablecaption{Vega to AB mag Conversion  \label{t_vegaab} }
\tablehead{
\colhead{Filter} & 
   \multicolumn{1}{c}{Vega - AB } 
}
\tablewidth{0pt}
\startdata
           GALEX FUV &  
                                         -2.223 \\
           GALEX NUV &  
                                         -1.699 \\
              SDSS u &  
                                         -0.944 \\
              SDSS g &  
                                          0.116 \\
              SDSS r & 
                                         -0.131 \\
              SDSS i & 
                                         -0.354 \\
              SDSS z & 
                                         -0.524 \\
   PanSTARRS gp1 & 
                                        0.103 \\
   PanSTARRS rp1 & 
                                       -0.133 \\
   PanSTARRS ip1 & 
                                       -0.361 \\
   PanSTARRS zp1 & 
                                       -0.516 \\
   PanSTARRS yp1 & 
                                       -0.543 \\
   PanSTARRS wp1 & 
                                       -0.084 \\
       Gaia DR2  G & -
                                        -0.092 \\
      Gaia DR2  G\_BP & 
                                        -0.066 \\
      Gaia DR2  G\_RP & 
                                        -0.362 \\
   Gaia DR2  G rev\tablenotemark{a} & 
                                       -0.079 \\
   Gaia DR2  G\_BP rev\tablenotemark{a} & 
                                       -0.073\\
   Gaia DR2  G\_RP rev\tablenotemark{a} & 
                                       -0.356 \\
\enddata
\tablenotetext{a}{For Gaia, the first set of values are computed with the passbands  
used in the Gaia DR2 database, the ``rev'' values are computed with the revised set of G, G\_BP and G\_RP  
						passbands based on a later 
						available nominal knowledge of the
						instrument, published by  \citet{evansgaiacalibation}. 
 }
\end{deluxetable}

\section{$GUVcat\_AIS$ matches with optical catalogs}
\label{s_match}

Below we describe the $GUVmatch$  catalogs, i.e.  $GUVcat$  matched with Gaia DR2 and SDSS DR14. We estimate the incidence of spurious matches relative to a given match separation, and discuss the fraction of multiple matches. The content of sources in the $GUVmatch$ catalogs in terms of broad astrophysical classes is discussed in  Section \ref{s_discussion}.   

\subsection{GUVcat\_AIS\_FOV055 match with Gaia DR2: $GUVmatch$\_AISxGaiaDR2} 
\label{s_gaia}

 The Gaia mission covers the whole sky, accumulating scans with a pattern that results in progressive completeness as the survey continues \citep{brownGaiadr1}.  We matched $GUVcat$\_AIS\_FOV055 \citep{bia17guvcat,bia20guvcat} with Gaia's data release 2 (DR2) \citep{gaiacoll18}.  A space-borne  instrument designed for precision astrometry, Gaia has much higher resolution than $GALEX$ and than the ground-based optical survey used here, but  some UV-bright objects detected by GALEX are below Gaia's detection limits in  Gaia's optical bands, and only about one third of the $GUVcat$ sources have a match in Gaia DR2. 

  Table \ref{t_statgaia}  provides a statistical overview to characterize the matched catalog at a glance. 
For the total matched catalog, and for  5{\mbox{$^{\circ}$}}~ Galactic latitude strips, we give 
the total number of matches, i.e. all Gaia DR2 distinct sources found within the 3{\mbox{$^{\prime\prime}$~}} match radius around the UV sources\footnote{in the matched catalog, a UV source with n matches is entered n times (one row for each distinct match); a unique list of sources can be immediately derived by selecting entries with tag $DISTANCERANK$ $<$2 (Table \ref{t_mmtags});  a list of sources with only one match can be extracted by selecting  $DISTANCERANK$ = 0}, the number of UV sources with Gaia counterparts, and of those
 with a unique Gaia match (tag $DISTANCERANK$ = 0), and the fraction of UV sources that have multiple Gaia matches.
The second row, for each Galactic latitude range, gives the  counts excluding sources that fall in the footprint of extended objects larger than 30{\mbox{$^{\prime}$~}}.  
 This overview is useful because one can examine characteristics (e.g., color distributions) of clean samples by selecting sources with a unique match, and then correct the resulting statistics for the discarded fraction of multiple-match sources. All  Gaia DR2 sources have a  flux measurement in at least the  $Gaia~G$~band, with additional measurements 
in $Gaia~BP$ and  $Gaia~RP$; not all Gaia DR2  sources have a significant parallax, since the [parallax~,~proper~motion] solution requires a sufficient number of repeated measurements and a successful fit to the data.  
In Table  \ref{t_statgaia} we  list also the number of Gaia matches that have a parallax measurement (excluding the parallaxes with negative values, that are a clear indication of solution failure), and of those that have parallax error $\leq$ 20/30/50\% (counted from the unique UV source list, column 3 of Table  \ref{t_statgaia}).  
For comparison, the number of UV $GUVcat$\_AIS sources and the area covered at different Galactic latitudes is given in Tables 6 and 7 of \citet{bia17guvcat}.
  Of the UV sources with Gaia matches, 1,977,196 
have significant FUV detection (with a total of 2,100,359 Gaia matches), of which  1,145,775/686,339/283,448  
 have errors $<$ 0.3/0.2/0.1~mag in both FUV and NUV. 

\begin{figure*}[!h]
\centerline{
\includegraphics[width=8.5cm]{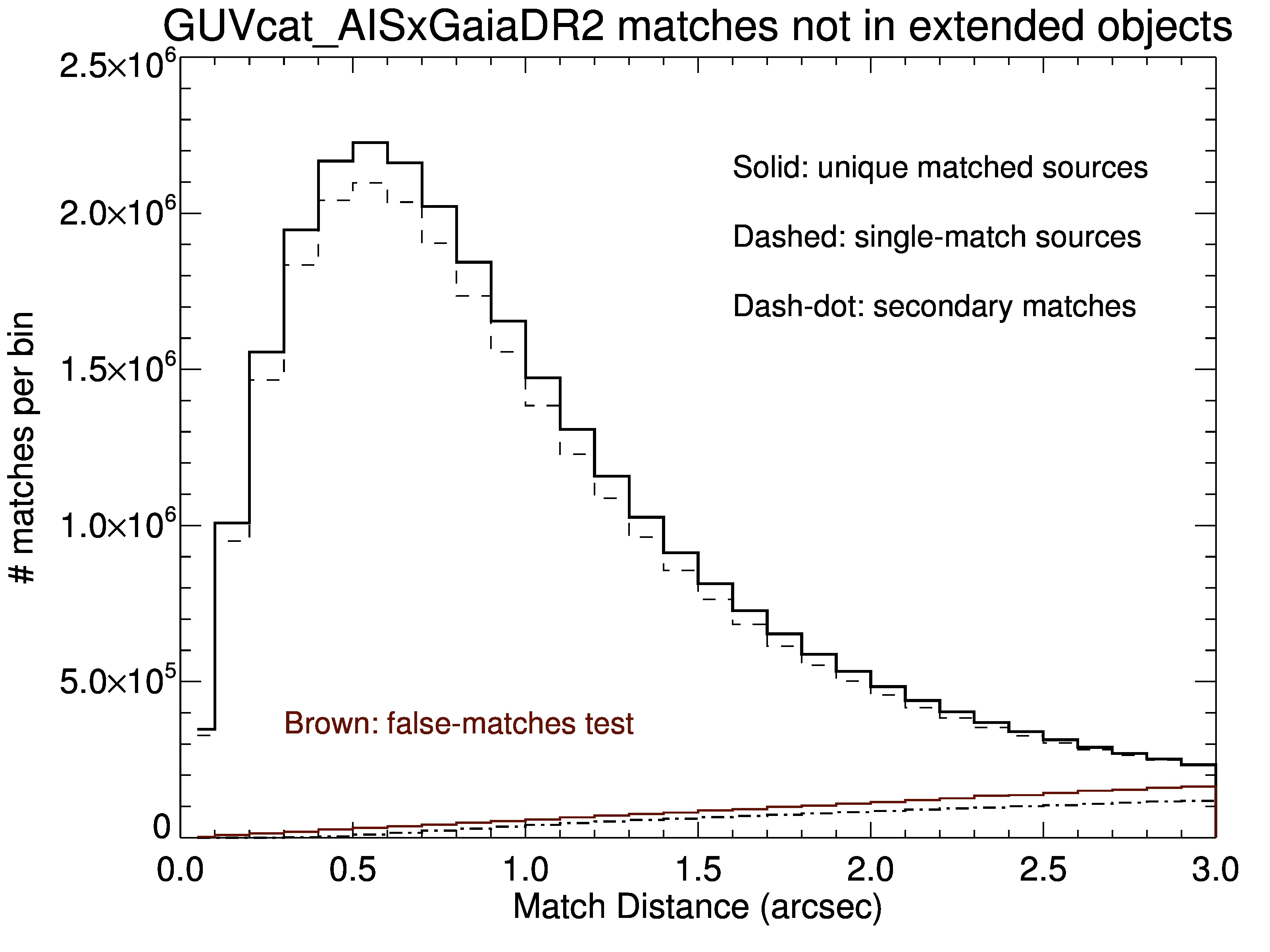}
\includegraphics[width=8.5cm]{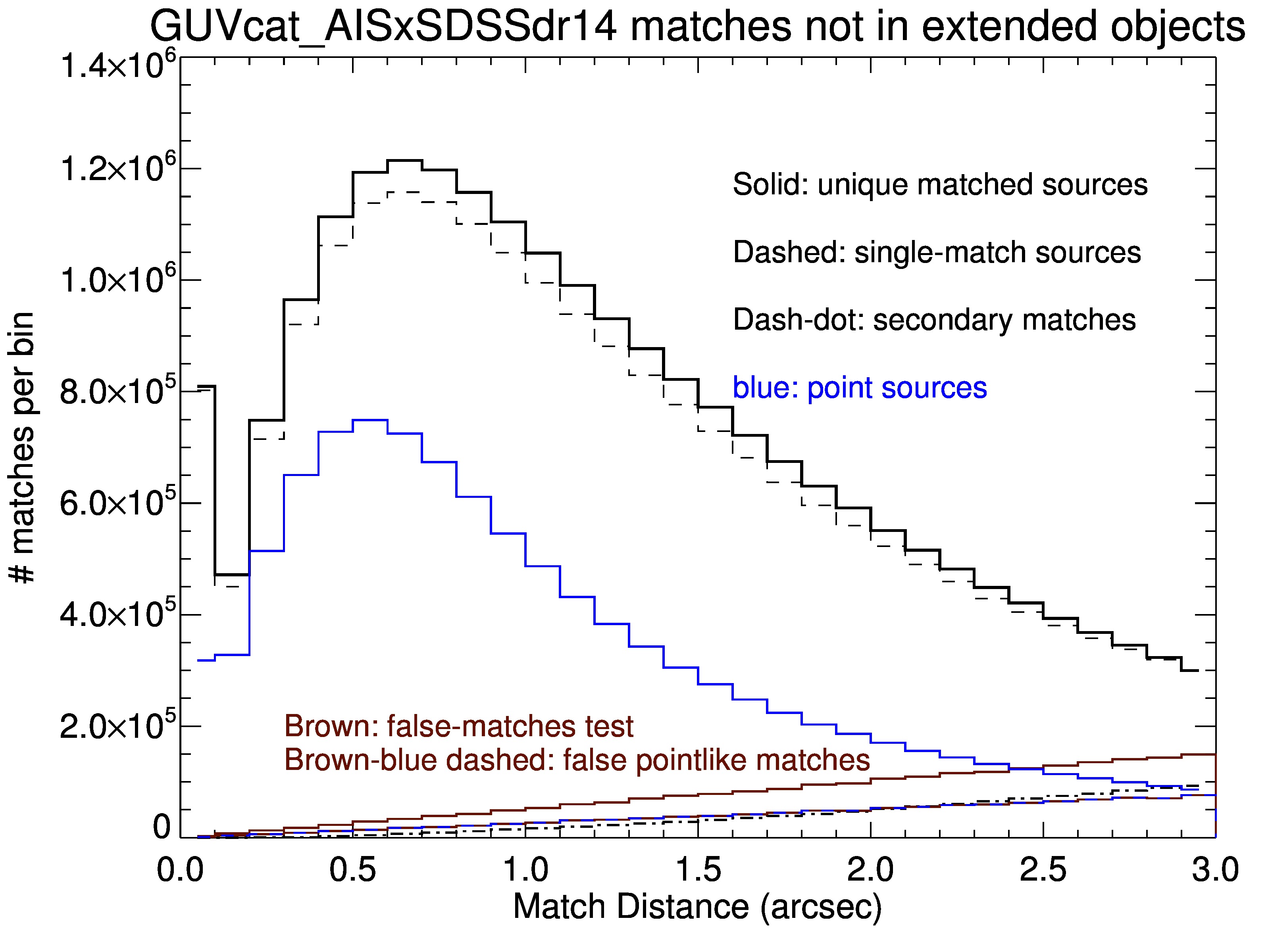}}
\caption{Distribution of separation between the $GUVcat\_AIS$ UV-source position and the  matched-source position (Gaia~DR2, left panel; SDSS DR14, right panel). 
 The distributions of the entire matched catalogs (counting the UV sources only once, solid line), and the sources with a unique match (dashed line) are qualitatively similar, because multiple matches are a small fraction (Section \ref{s_intro}, Tables \ref{t_statgaia} and \ref{t_statsdss}). The separation distribution of primary matches peaks at values around 0.6{\mbox{$^{\prime\prime}$~}}, while that of the secondary matches (dash-dot line) increases towards larger separations. 
 For the SDSS matches, point-like sources are also shown separately (blue histogram), since extended sources may have a looser definition of source center across the wavelengths. To assess the fraction of spurious matches, we matched each optical database with a replica of $GUVcat\_AIS$ with altered coordinates; the distribution of false matches is shown in brown.  In the counts, we excluded sources in extended objects such as galaxies or stellar clusters larger than 30arcmin (see \citet{bia17guvcat}). 
\label{f_matchdist} }
\end{figure*}

\begin{figure*}[!h]
\centerline{
\includegraphics[width=13.cm]{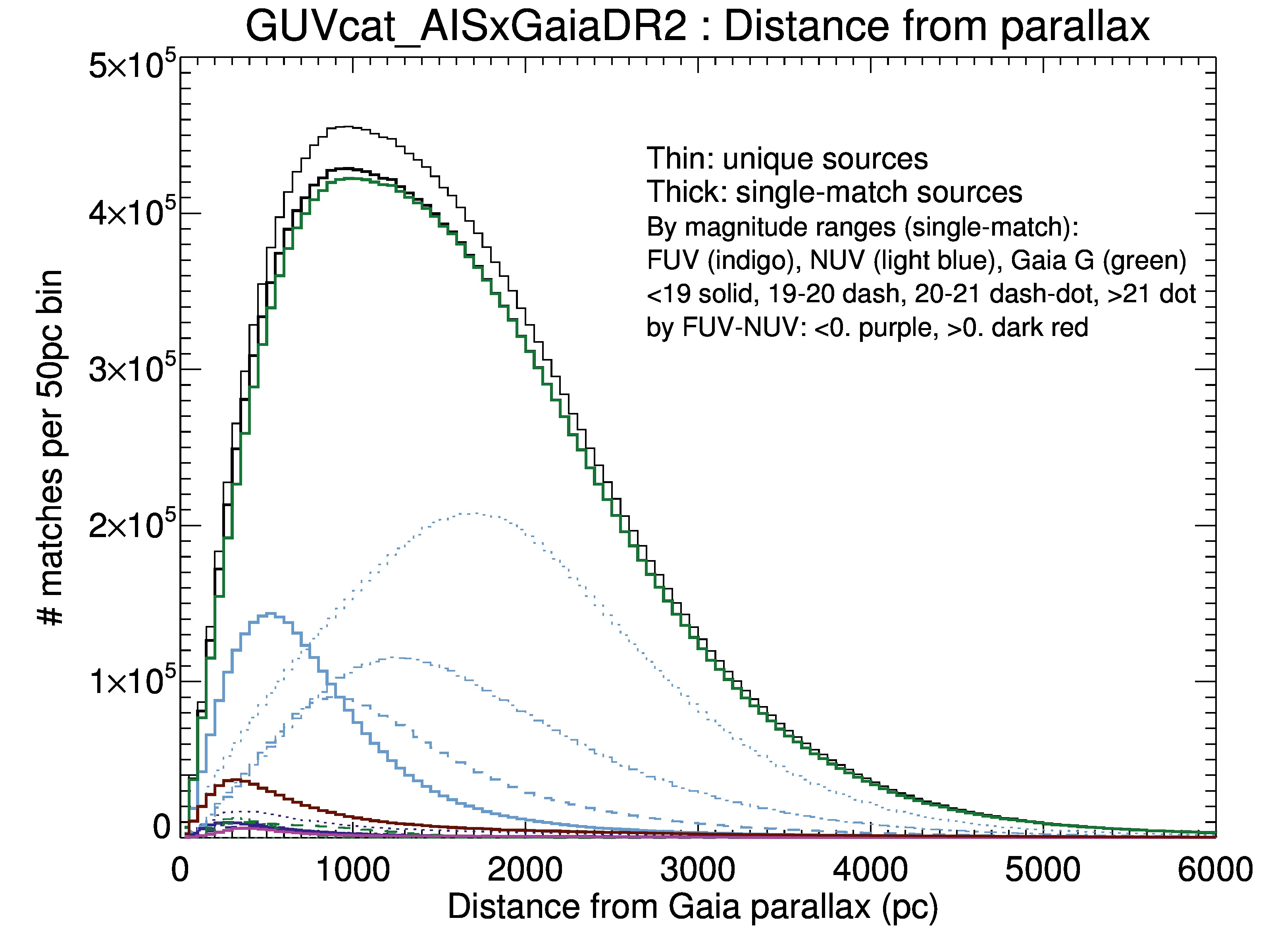}
}
\vskip -0.3cm
\centerline{
\includegraphics[width=13.cm]{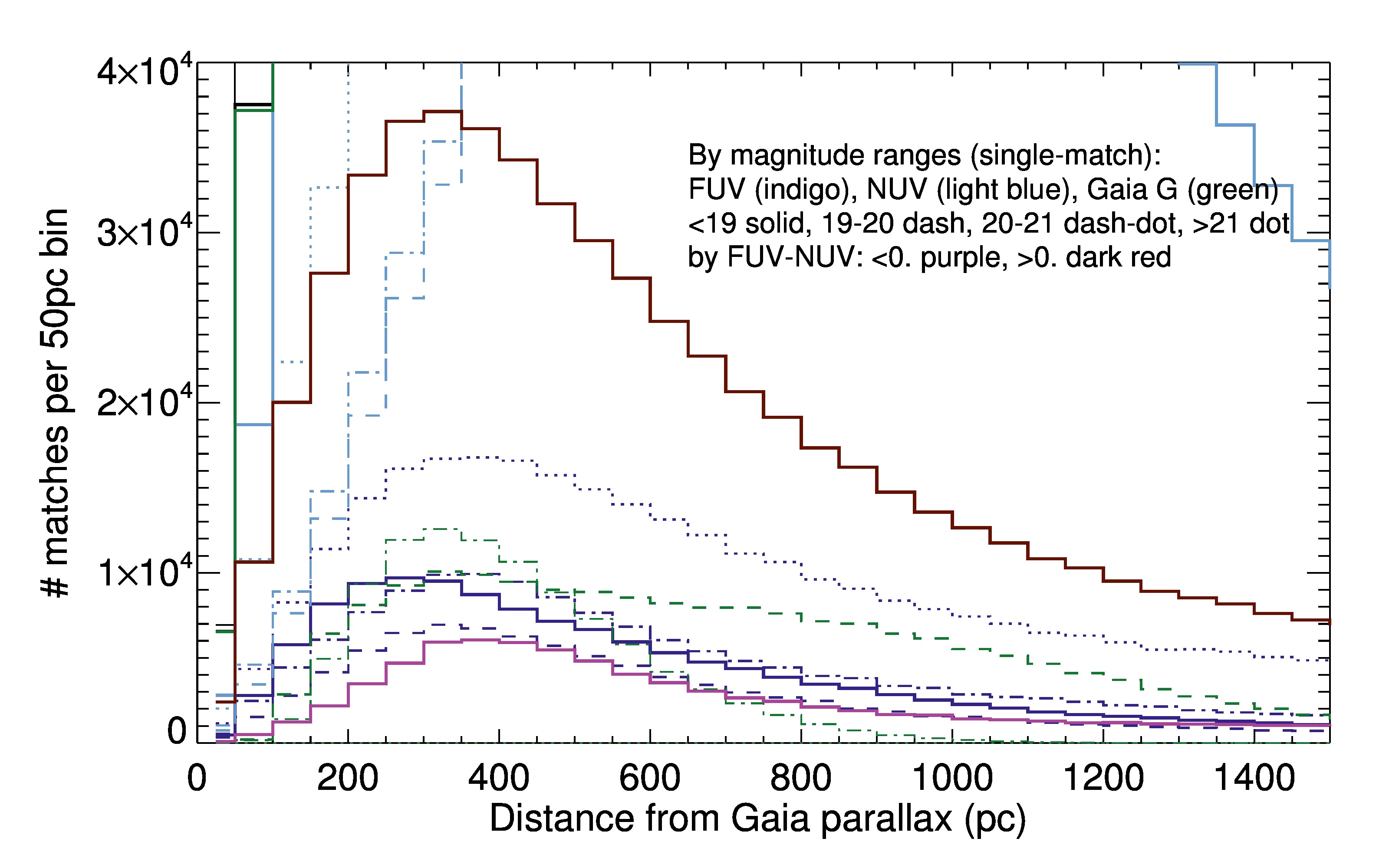}
}
\caption{Distribution of distances (pc) derived from the Gaia parallax for the matched sources with parallax error less than 50\%. 
The bottom panel is a zoom to distances $<$1500pc. 
  Sources with FUV detection are about $\sim$10\% of the NUV detections (see text) and are distributed over a range of magnitudes; Gaia parallax measurements are more numerous for sources with FUV-NUV$>$0, that are relatively brighter at optical wavelengths than the hotter stars.  
The plot also illustrates the unique sensitivity of UV surveys to hot WDs, with respect to optical surveys (e.g., \citet{bia11a,bia18b}), and why Gaia's WD sample hardly extends beyond a few hundred parsecs \citep{gaiababu18} ). Sources in objects larger than 30{\mbox{$^{\prime}$~}} have been excluded, as well as those with negative parallax values. 
 \label{f_pardist} }
\end{figure*}

 Figure \ref{f_matchdist} shows the distribution of matched-sources separation, i.e. the difference between the $GUVcat$ and Gaia DR2 position (tag $DSTARCSEC$ in the matched catalog).  For nearby objects, if they have a high proper motion, part of the difference may also be ascribed to actual source motion in the plane of the sky\footnote{
the measurements in $GUVcat$ may have been taken at any time during  GALEX operations with both detectors; the observing date of each entry can be found by searching for the ``OBJID'' identifier  in the Casjobs Visit-level (if ``CORV'' = V) or MCAT-level (if if ``CORV'' = C) tables, and retrieving the additional tags of that measurement; GALEX coordinates are given at a reference epoch of 2000., equinox = 2000. 
Gaia DR2 (\url{https://www.cosmos.esa.int/web/gaia/dr2}) is built from data collected  between 25 July 2014 and 23 May 2016; the reference epoch is J2015.5.}, although most UV GALEX sources are probably rather distant (see e.g., \citet{bia11a,bia18b} and Figure \ref{f_pardist}).   Gaia, given its spatial resolution, and its main objective and corresponding mission design, has a superior astrometry with respect to GALEX; however, an important caveat must be recalled concerning sources in close binaries. Whether real binaries or chance alignments along the line of sight, Gaia's pipeline - in DR2 -  does not solve for pairs closer than 2{\mbox{$^{\prime\prime}$~}}. According to the current documentation, if two sources closer than 2{\mbox{$^{\prime\prime}$~}} are detected, only one is kept, but 
not necessarily the same one throughout the repeated  observations, 
 potentially hampering the [parallax, proper motion] solution.  This caveat affects probably a very small number of sources, especially considering that our starting point is a catalog of UV sources, rather sparse in comparison to the density of optical sources. It is, however, relevant for some specific science goals, such as for example our study of hot white dwarfs in binaries (e.g., \citet{bia11a,bia18b}, see Section \ref{s_discussion}).  As recalled in Section \ref{s_introguvcat}, the centroid position of $GALEX$ sources is as good as 0.32-0.35/0.34-0.48{\mbox{$^{\prime\prime}$~}} (NUV/FUV) in GR6/7, except possibly in the periphery of the fields. Therefore, if we were to restrict the sample to matches within a smaller match radius (as can be easily accomplished by applying a restriction in tag $DSTARCSEC$ in our catalog), we might gain by eliminating some spurious secondary matches, but at the price of possibly losing binaries and high proper motion objects.

 To estimate the incidence of spurious matches, i.e. accidental positional coincidences, we performed a test by matching Gaia DR2 with a fake $GUVcat$-like catalog, made by picking one every ten $GUVcat$ sources, and offsetting their position by 5arcmin.  The $\gtrsim$8~million sources ``fake-$GUVcat$'' catalog has 252,135 Gaia DR2 matches; their distribution with separation (difference between the position of the GALEX source and that of the Gaia match) is also shown in Figure \ref{f_matchdist}~(brown histogram). For a consistent comparison with the actual catalog, the result of the ``fake-$GUVcat$'' match is multiplied by 10, because we used a one-tenth fake $GUVcat$ catalog.   The number of matches  per bin increases with separation because the annuli with increasing radius cover a progressively larger area; we do not expect random matches to favour small separations. Figure \ref{f_matchdist} shows that at a separation of 3{\mbox{$^{\prime\prime}$~}} (the chosen match radius),  the number of spurious matches almost reaches  the number of matches to the real $GUVcat$ positions. Therefore, we can consider our catalog rather complete (no real counterparts beyond the matched radius, for point-like sources). Figure \ref{f_matchdist} also shows that the secondary matches (the additional matches with  separation larger than that of the closest match) could be mostly random alignments; but even though the closest match may be the [only] real counterpart,  the UV flux can be composite of the primary and secondary match, given the GALEX resolution,  as discussed earlier.  Therefore, sources with multiple matches should be examined with care. 

 The Gaia DR2 database gives Gaia  magnitudes in the Vega~magnitude system.  We kept the original magnitude values in $GUVmatch$\_AISxGaiaDR2. 
Therefore, in $GUVmatch$\_AISxGaiaDR2,  GALEX's FUV and NUV magnitudes are AB mags, and Gaia G, BP, RP are Vega~mags; for SED analysis they must be converted to one of the two systems, by applying to one of the two sets the transformation coefficients taken from Table \ref{t_vegaab} or other sources. 
 We provide in Table \ref{t_vegaab} the transformation coefficients between Vega and AB magnitude systems, calculated using the Vega spectrum and the transmission curves for all filters included in our catalogs. For Gaia DR2, we calculated the coefficients with  the transmission curves used by the Gaia project for the DR2 photometry released in 2018, and also with the revised curves published by  \citet{evansgaiacalibation}.  Other revisions are published, as inflight instrumental response functions continue to be investigated and refined; an updated summary is given in the Gaia documentation pages,  \url{https://www.cosmos.esa.int/web/gaia/dr2-known-issues\#PhotometrySystematicEffectsAndResponseCurves}. 
  Our computed Gaia conversion coefficients are not identical to the zero-point differences reported in the Gaia online documentation\footnote{\url{https://gea.esac.esa.int/archive/documentation/GDR2/Data\_processing/chap\_cu5pho/sec\_cu5pho\_calibr/ssec\_cu5pho\_calibr\_extern.html}, see Tables 5.2 and 5.3 in that section.}, the differences are up to $\approx$4 hundredths of a magnitude, and are probably due to the interpolation of the transmission curves downloaded and the adopted Vega spectrum.

Figure \ref{f_pardist} shows the distribution of distances, derived from Gaia parallax\footnote{we added to the parallax a bias of 0.052~mas, following \citet{gaiabias}}, for the matched Gaia DR2 UV sources  that have parallax error less than 50\%. 
Because the sources with FUV detection are only about one tenth of the NUV detections, we show 
also an expanded view of the FUV-detected sources  within 1500~pc.  We also separate the sample in ``hotter'' and ``cooler'' sources, according to their  FUV-NUV color of $<$0 or $>$0 respectively.  For stars, the GALEX FUV-NUV color is a good indicator of stellar temperature,  since it is almost reddening-free for typical Milky Way (MW) dust with {$R_V$~}=3.1 (\citet{bia17guvcat}, their Table 1, see also Section \ref{s_discussion}).  The cut at FUV-NUV=0 separates stars hotter than $\sim$20,000~K (the exact value depending on gravity and metallicity) from cooler stars and from the majority of extra-Galactic sources.    To separate single hot stars from binaries, which is one of our motivations for building these matched catalogs, and cooler stars from extra-Galactic sources, different color combinations are useful, as shown in Section \ref{s_discussion} (see also e.g., \citet{bia09,bia11a} for additional discussion).
 
A higher fraction of the  ``cooler'' UV sources have Gaia parallaxes than the hotter ones, as can be expected because very hot stars are optically fainter than cooler stars with similar UV brightness, as will be  illustrated in Section \ref{s_discussion}.  
We point out that, because the FUV-NUV color requires a detection in both filters, and the fraction of sources with FUV detections is about one tenth of the $GUVcat$ total NUV sample overall,  the fraction of ``hotter'' and ``cooler'' stars shown in Figure \ref{f_pardist} is a small subset of the total $GUVmatch$\_AISxGaiaDR2 sample; the number of sources with good parallax measurements increases further for NUV sources undetected in FUV, that are cooler yet.

\subsection{GUVcat\_AIS\_FOV055  match with SDSS DR14:  $GUVmatch$\_AISxSDSSdr14} 
\label{s_sdss}

The Sloan Digital Sky Survey (SDSS, \citet{sdss}) provides five-band {\it u, g, r, i, z } photometry over a large part of the northern sky. The overlap with the $GUVcat$\_AIS footprint is shown in Figure 1 of \citet{bia14uvsky} in Galactic coordinates. The figure is not repeated here as the difference with respect to earlier releases will not be appreciable. The total current overlap is $\sim$11,100~square degrees (Table 1 of \citet{bia19areacat}); the area of overlap in any chosen region of the sky can be calculated with the online tool $AREAcat$\footnote{\url{http://dolomiti.pha.jhu.edu/uvsky/area/AREAcat.php} }  \citep{bia19areacat}.    The match with SDSS data release 14 (DR14\footnote{\url{https://www.sdss.org/dr14/}}, \citet{sdssdr14}) yields  23,310,532 
 counterparts within 3{\mbox{$^{\prime\prime}$~}} of 22,207,563  
unique $GUVcat$\_AIS sources, 10,167,460 
of which are point-like according to the SDSS pipeline photometry.  SDSS spectra exist for 860,224 
matched UV sources.  In the SDSS DR14 database, some sources with the same photometric identifier are repeated because  two spectra exist; there are 21 such repeats in the resulting matches, for these we kept only one entry, for the sake of counting how many unique $GUVcat$ sources have matches.  Therefore, only one of the available SDSS spectra is listed for these cases. $GUVcat$ starting point is a list of all sources detected in NUV, regardless of FUV detection;  among the matched sources, 3,511,890 
have also  significant detection in GALEX-FUV, of these  1126671 / 427361 / 103077 
have error less than 0.3/0.2/0.1~mag in both FUV and NUV. 

 More statistical figures are  given in Table \ref{t_statsdss}, reported for the entire matched database and by Galactic latitudes.\footnote{the area of overlap between $GUVcat$\_AIS and SDSS DR14 by slices of Galactic latitude can again be found in Table 1 of \citet{bia19areacat}, or calculated with the $AREAcat$ tool for any desired region (\url{http://dolomiti.pha.jhu.edu/uvsky/area/AREAcat.php}).} As in Table \ref{t_statgaia}, the second row for each Galactic latitude range gives the source counts, excluding sources that fall in the footprint of extended objects larger than 30{\mbox{$^{\prime}$~}}. 
Due to the scant sky coverage of the UV surveys near the Galactic plane, and the significant drop in UV source measurements at very low Galactic latitudes because of dust extinction, as shown dramatically in Figure 2 of \citet{bia11a},  and to  the SDSS footprint favouring the Northern latitudes (Figure 1 of  \citet{bia14uvsky}), the number of SDSS~DR14  matched sources significantly decreases at low Galactic latitudes, although the fraction of multiple matches increases, owing to the actual higher density of stars in the Milky Way  disk. This is shown in Figure 3 of \citet{bia11a} for an earlier version of the matched catalog; although the present matched catalog is significantly expanded both in coverage and quality  with respect to the 2011 version, and includes additional science-facilitating tags,  we do not repeat the figure here because it would be very  similar. 

\begin{figure*}[!h]
\centerline{
\includegraphics[width=16.cm]{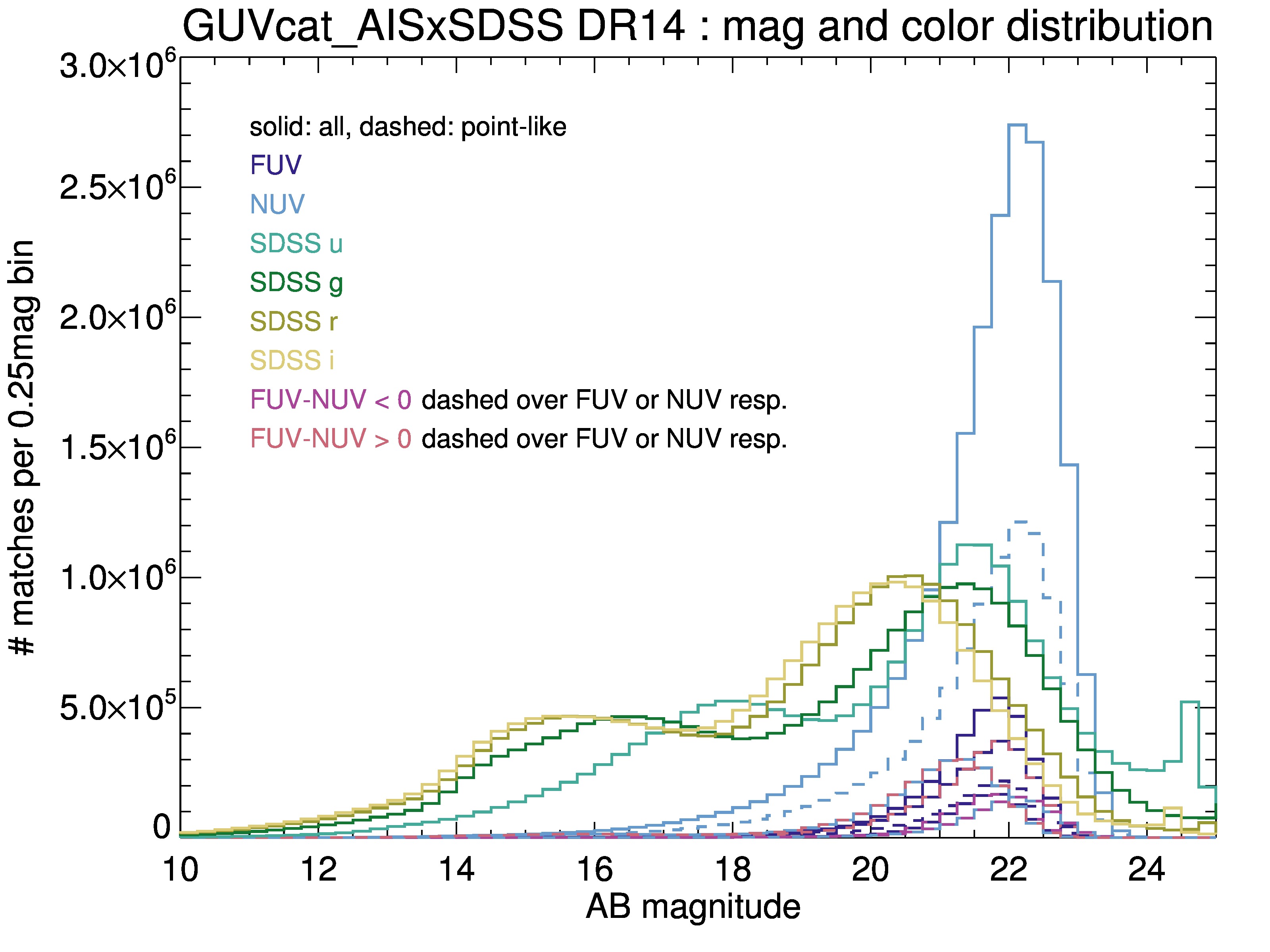}
}
\caption{Distribution in FUV, NUV, {\it u, g, r, i} magnitudes of the matched $GUVcat$\_AIS~X~SDSS~DR14 sources, with color coding indicated in the legend. Dashed lines are for point-like sources, classified as such from SDSS imaging.  We also plot FUV and NUV magnitude histograms of sources separated by FUV-NUV $>$0  and FUV-NUV $<$0,  with dashed  light-red and fuchsia respectively, over indigo (FUV) and light-blue (NUV)  histograms.  Sources in objects larger than 30{\mbox{$^{\prime}$~}} have been excluded. The bimodal distribution in optical passbands is mainly due to the presence of both Galactic and extra-Galactic populations, as seen in Figures \ref{f_colcolsdss1} and \ref{f_colcolsdss2}.   
 \label{f_maghistsdss} }
\end{figure*}

Figure \ref{f_maghistsdss} shows the distribution  of $GUVmatch$\_AISxSDSSdr14 sources in FUV, NUV, {\it u, g, r, i } magnitudes, color-coded by filter. As in Figure \ref{f_pardist}, for a cleaner sample, we excluded sources in extended objects larger than 30{\mbox{$^{\prime}$~}};  these are a very small fraction (Table \ref{t_statsdss})  and do not change appreciably the histograms.  For FUV and NUV, we also plot separately   the counts for sources that have ``point-like'' morphology according to the SDSS pipeline (dashed-line histograms).  As for the Gaia matched catalog (previous section), we also plot counts separating sources by FUV-NUV color; these are shown with pale red (FUV-NUV $>$0 ) or fuchsia (FUV-NUV$<$0) dashes mixed with indigo (for FUV) or  light blue (for NUV) dashes. The sample with   FUV-NUV $>$0 includes stars cooler than $\sim$ 20,000~K, QSOs, and galaxies, as shown in Figures \ref{f_colcolsdss1} and \ref{f_colcolsdss2}, and discussed by \citet{bia11a}.  Stars and QSOs are classified morphologically as ``point-sources'' (type=STAR) in the SDSS database, while most galaxies are classified as extended (type=GALAXY), as shown in Section \ref{s_discussion}. Sources with  FUV-NUV$<$0 are mostly hot stars, hot WDs in particular, except for some low-redshift QSOs with enhanced {\mbox{Ly$_{\alpha}$}}
  The bimodal distribution of sources in optical magnitudes mainly reflects the separation between stellar and  extra-Galactic objects, as discussed by \citet{bia11a,bia11b} and shown in Section \ref{s_discussion}. 
In spite of the SDSS DR14 magnitude depth\footnote{According to the online documentation for DR14 ( https://www.sdss.org/dr14/imaging/other\_info/ ),
 the median 5-${\sigma}$ depth for SDSS photometric observations, based on the formal errors from PSF photometry on point sources, is: 
$u$ = 22.15,
$g$ = 23.13, 
$r$ = 22.70, 
$i$ = 22.20, and 
$z$ = 20.71 ABmag}
being $\approx$2-3mag fainter than $GUVcat$\_AIS,  GALEX can detect the hottest stars (WDs) at further distances and with smaller radii than SDSS (\citet{bia11a} and Section \ref{s_discussion}).

\section{Discussion. The Content of Astrophysical Sources in the Matched Catalogs.} 
\label{s_discussion}

To examine the content of different classes of astrophysical sources in the matched catalogs, we show color-color plots of the matched sources for representative samples at North Galactic latitudes, that are well covered by both Gaia and SDSS, and by $GUVcat$\_AIS, in Figures \ref{f_colcolsdss1} to \ref{f_colcolgaia2}. Such plots allow us to separate different astrophysical objects, regardless of their distance, that is known only for the nearest sources, and would anyway introduce an additional uncertainty, not necessary (initially) for the purpose of broad  source classification.   We follow, as a starting point,  the basic color-color plots used in our earlier studies \citep{bia07apj, bia11a, bia11b, bia09}, with additional models and details.  We  show color-color combinations, out of several possible, that best illustrate how and when different colors enable a clear separation (or not) of certain classes of sources, and of certain parameters ranges within these classes. These plots are an important first  guide to possible uses of our catalogs, because, in extracting  samples of selected candidate sources,  it is also necessary  to evaluate what other objects may intrude the selection, to estimate the purity of the sample, or - in other terms - the probability that the selected candidates belong to the intended class. 
 Examples of scientific applications are the identification of Milky Way's  hot white dwarfs (WD), and in particular hot WDs with a less evolved (cooler, optically brighter) companion, by \citet{bia11a,bia18b} and Bianchi et al. (in preparation). \citet{bia11a} also published a quantitative comparison of hot-WD counts at different Galactic latitudes with Milky Way models, and showed that the selection from the GALEX database reaches (optically elusive) hot WDs out to the Milky Way halo, and that extracted samples are sensitive to critical and yet poorly constrained aspects of stellar evolution, such as the Initial-Final Mass Relation (IFMR), that maps the final WD mass to the initial mass of its progenitor.  Early model comparisons by \citet{bia11a} favoured IFMR with lower final masses, such as \citet{weidemann2000} IFMR. More recent results \citep{bia18c} hinted that the binary fraction of  observed WD samples is 
 lower than that reported for unevolved stars, implying the possibility of a significant merging rate during binary evolution. But model-predicted source counts are sensitive to several ingredients, including assumptions on stellar evolution  and on the geometry of stellar populations and interstellar dust (that affects observed magnitude distributions and stellar counts to given brightness limits) in the Milky Way. Comprehensive Milky Way model comparisons of the stellar samples in our matched catalogs  are under way (dal~Tio et al in preparation, Bianchi et al. in preparation, Million et al. in preparation). 

Figures \ref{f_colcolsdss1} to \ref{f_colcolgaia2} illustrate the power  of SED measurements  extending to  UV wavelengths for identifying and broadly classifying types of  astrophysical sources that are either elusive or poorly characterized at longer wavelengths.  In particular we recall that, for Milky~Way-type extinction with {$R_V$~}=3.1, the GALEX FUV-NUV color is almost reddening-free \citep{bia11,bia17guvcat}, as also shown by the reddening arrows drawn on the color-color plots in Figures \ref{f_colcolsdss2},  \ref{f_colcolgaia2} and \ref{f_colcolgaia3}. Therefore,  FUV-NUV is a direct, robust indicator of stellar {\mbox{$T_{\rm eff}$}~} (for {$R_V$~}=3.1-type extinction), while UV-optical or optical-only colors are highly affected by reddening, that must be accounted for in deriving stellar parameters from the observed SED, and carries an additional uncertainty.   In addition, the plots illustrate the well-known greater sensitivity of UV data to the hottest {\mbox{$T_{\rm eff}$}}s,    to which optical colors are saturated, and to some other sources. 

\subsection{Point-like and extended sources}
\label{s_pntext}

To separate point-like and  extended sources in the matched catalogs,  let's first recall that  SDSS imaging has a $\sim$3$\times$ higher resolution than GALEX imaging. Therefore, for $GUVmatch$\_AISxSDSSdr14 sources, we use the SDSS tag ``type'', which is in fact a classification based on the source morphology. A {\it type}=STAR source means point-like (at the SDSS resolution), and includes all stars and most if not all  QSOs,  while  {\it type}=GALAXY selects  extended sources, that are mostly galaxies, as shown in Figures    \ref{f_colcolsdss1} to  \ref{f_colcolsdss2}. 

Gaia DR2 does not include a similar morphological classification parameter. In DR2, photometry of extended sources is not specifically measured. Extended sources above the brightness limit are, however, included in the DR2 database; therefore, when measured with the same treatment as point sources, they result in a poor fit to the Gaia  point-spread function (PSF).  A measure of the fit quality, i.e., of how much a source departs from, or matches, the Gaia PSF, is given by  the tags  {\it astrometric\_excess\_noise} 
and {\it phot\_bp\_rp\_excess\_factor}
 (see e.g., \citet{evansgaiacalibation}).  
 We mark with green dots in Figures \ref{f_colcolgaia1} to \ref{f_colcolgaia3} the Gaia sources with
{\it  astrometric\_excess\_noise} $>$ 5 and {\it phot\_bp\_rp\_excess\_factor} $>$2, following Cheng et al. (private communication). The green dots cover sources in the galaxies' color locus [NUV-G, BP-RP] (by analogy with the UV-SDSS color-color diagrams, and with model loci), and possibly also include some binaries (Section \ref{s_bina}). In the NUV-G vs FUV-NUV diagrams, however, they are mostly ``bluer'' than the stellar sequence in  NUV-G (not so in the NUV - SDSS-r color).  We used in these plots the GALEX ``best'' magnitudes, that is, the measurements deemed by the GALEX pipeline to be the most adequate for each source. For extended sources, these are likely not aperture magnitudes. 
In the case of an extended source, GALEX and SDSS pipelines integrate over the entire object, while Gaia DR2  performs only a PSF fit, making GALEX-Gaia colors possibly inconsistent in such cases. Our catalog  also includes GALEX aperture magnitudes with varying aperture sizes; in the color-color plots we used the ``best'' magnitude throughout the sample for consistency.  This criterion to identify extended sources in Gaia DR2 is explored here for the purpose of eliminating potentially extended sources from stellar samples.

\subsection{Separating stars, QSOs, and galaxies}
\label{s_colcol}

\begin{figure*}[!h]
\vskip -3.5cm
\centerline{
\includegraphics[width=18.cm]{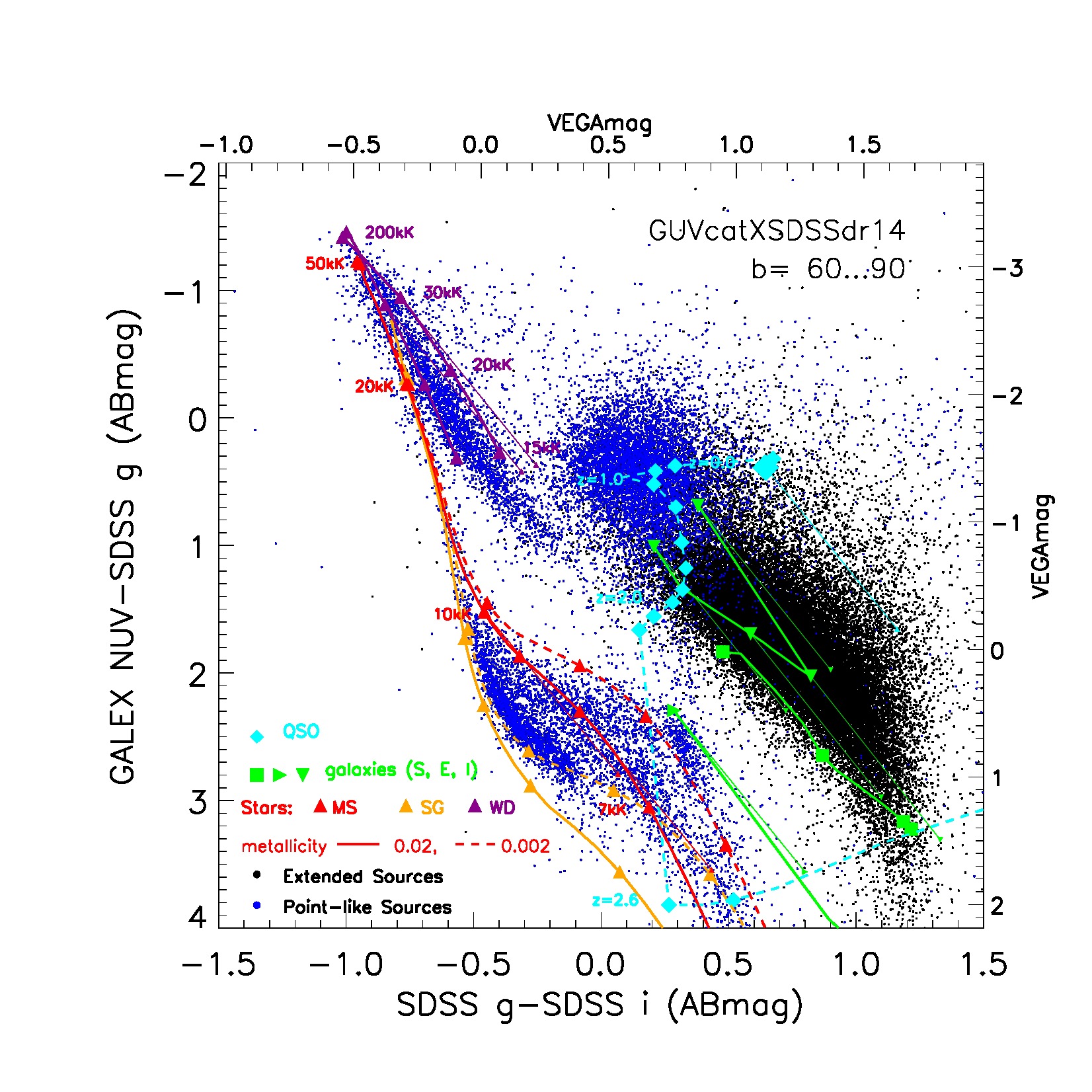} 
}
\vskip -1.5cm
\caption{Color-color plot of 
$GUVmatch$\_AISxSDSSdr14 sources in the 60-90{\mbox{$^{\circ}$}} North Galactic latitude range (dots; blue for  point-like sources). Lines show model colors for stars of varying {\mbox{$T_{\rm eff}$}~}'s (symbols mark {\mbox{$T_{\rm eff}$}~}=50, 20, 10, 9, 8 and 7~kK), galaxies of varying ages, and QSOs (redshift=0 (large cyan diamond), .2, .4, .6, 1.0, 1.2, 1.4, 1.8, 2.0, 2.2, 2.4, 2.6, 3.0, and 4.0), see legend and Section \ref{s_discussion}.  Reddening vectors for {\mbox{$E_{B\!-\!V}$}}=0.3mag are shown as thin arrows starting from some unreddened model colors. 
 Two purple {\mbox{$T_{\rm eff}$}~} sequences ({\mbox{$T_{\rm eff}$}~}=200kK to 15kK) are pure-H WD colors for {log~{\it g}}=7 and 9. Numerous point-like sources within these sequences identify the Milky Way's hot WDs \citep{bia11a}.   Point-like sources occupy the loci of stellar and QSO models, while galaxy model colors account for most extended sources. Other conspicuous stellar densities are seen for low-metallicity supergiants,  detected out to large distances towards high Galactic latitudes, and main sequence stars of solar and sub-solar metallicity, that may constitute the thin and thick disk contributions.  The actual number of stars cooler than $\sim$8,000K steeply increases, but cooler sources, fainter in UV bands, are eliminated in the plot by error cuts (that translate into brightness cuts).  Low-z QSO models with enhanced {\mbox{Ly$_{\alpha}$}}
cover the extent of the round lump of point-like sources approximately centered on redshift=1 QSOs. 
\label{f_colcolsdss1}
}
\end{figure*}

\begin{figure*}[!h]
\vskip -3.cm
\centerline{
\includegraphics[width=9.5cm]{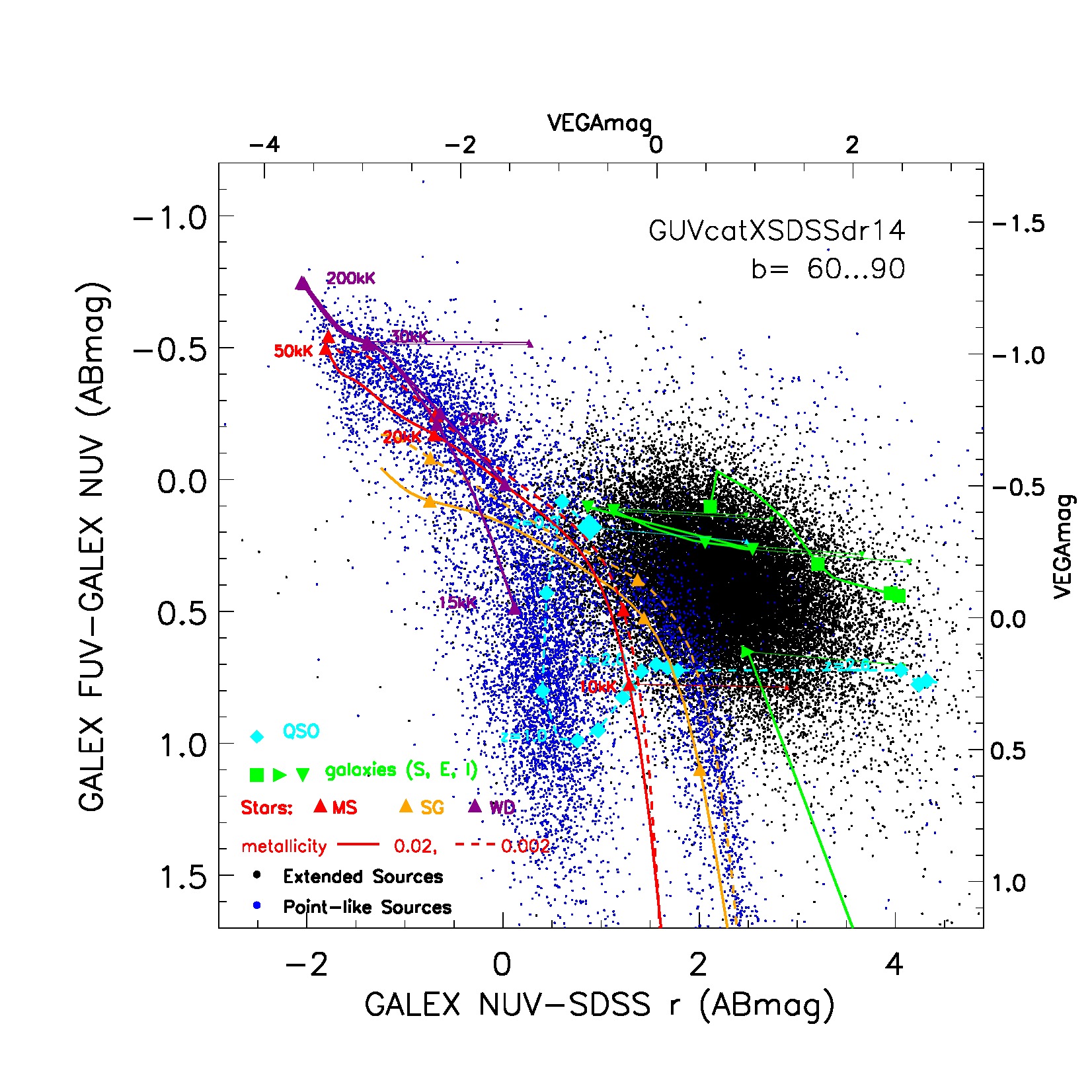}
\includegraphics[width=9.5cm]{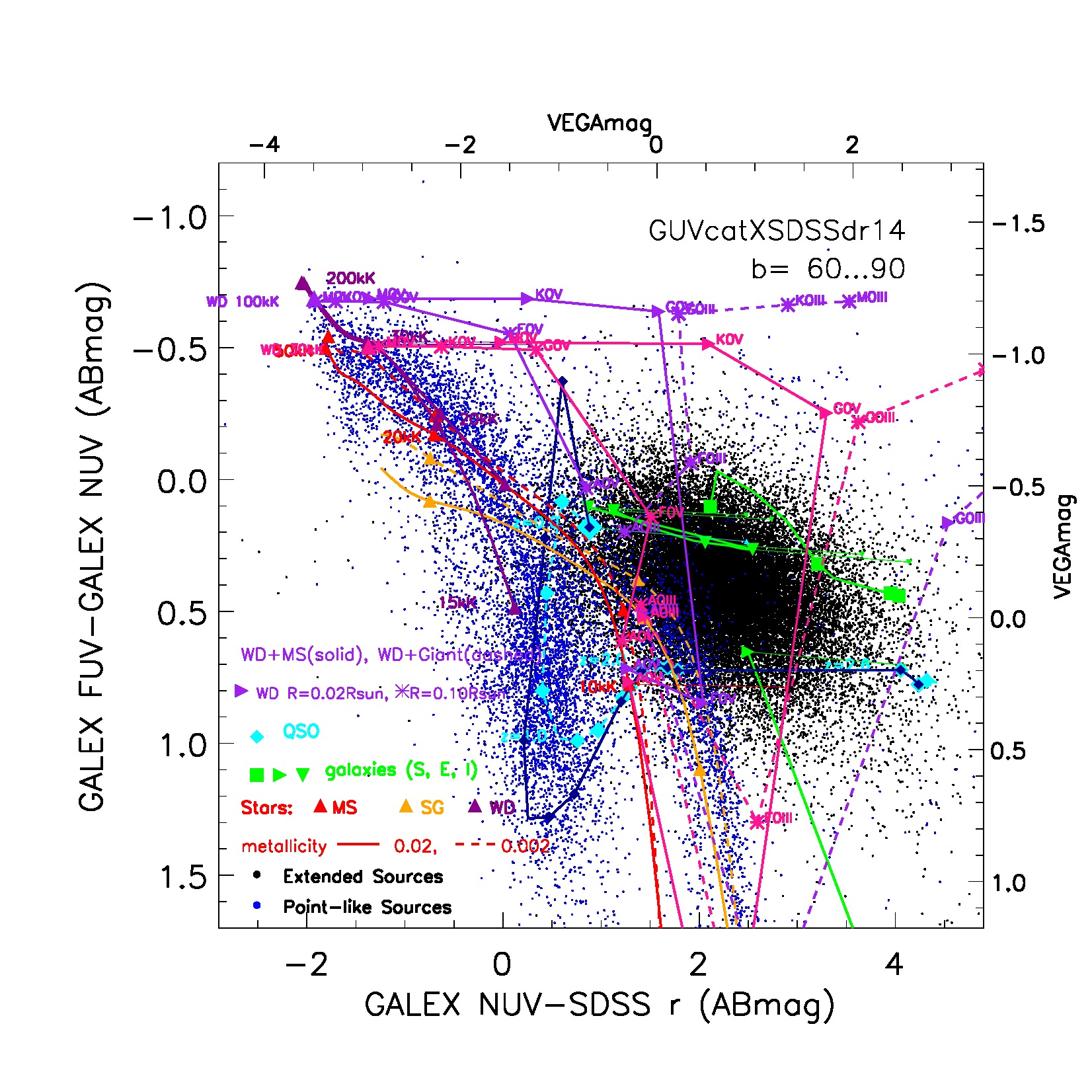}
}
\vskip -0.5cm
\centerline{
\includegraphics[width=9.5cm]{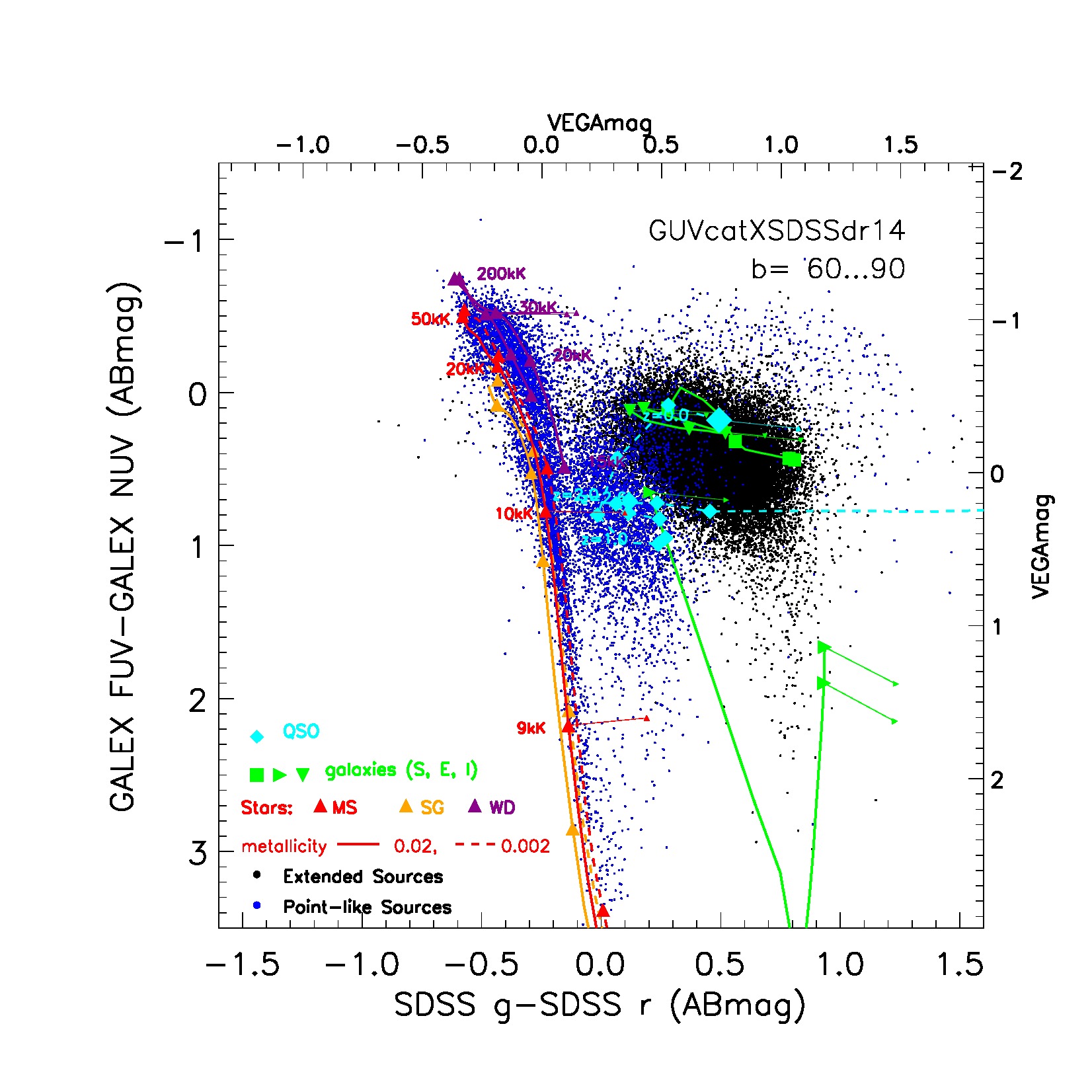}
\includegraphics[width=9.5cm]{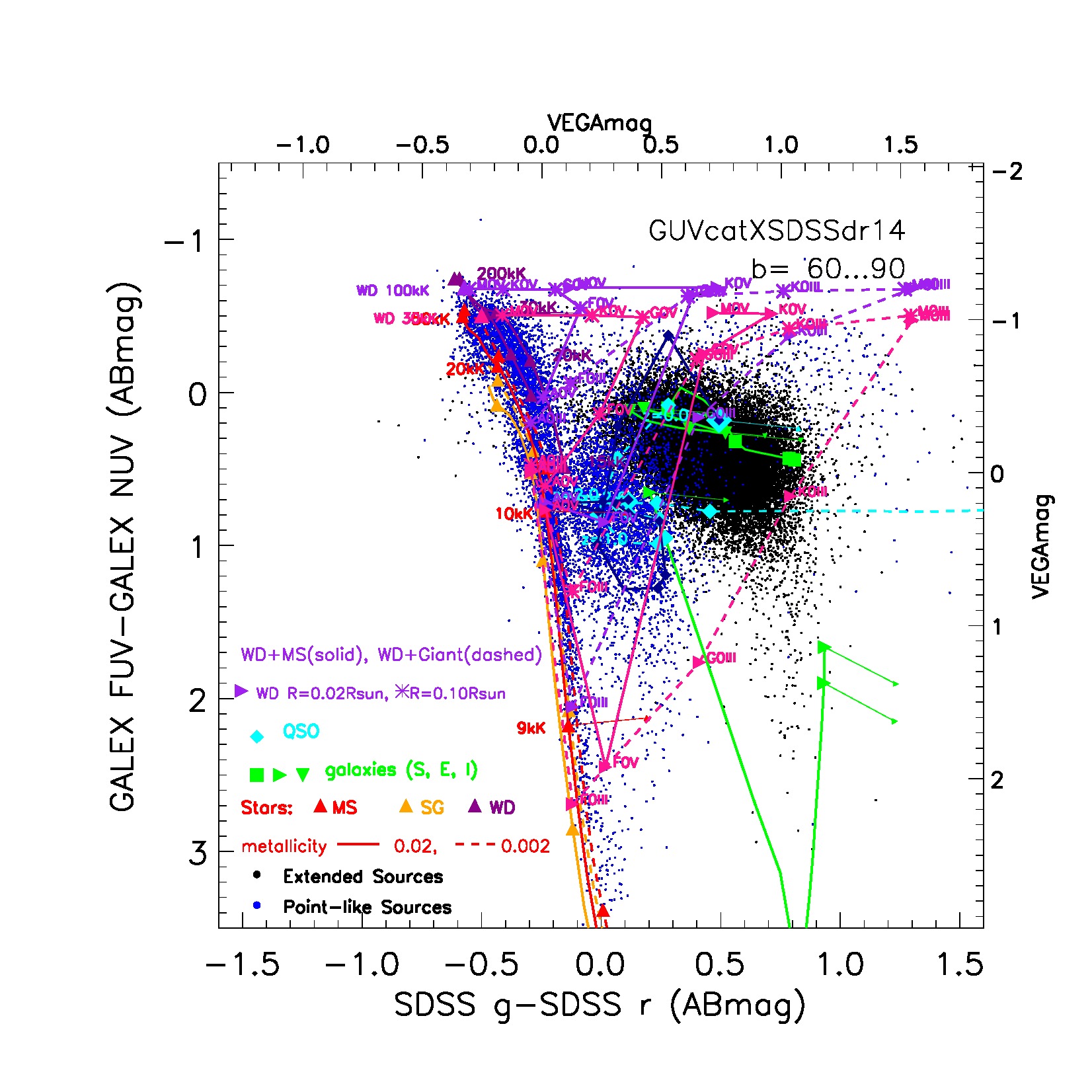}
}
\vskip -1.cm
\caption{Color-color plots of $GUVmatch$\_AISxSDSSdr14  sources,  in the  60-90{\mbox{$^{\circ}$}} N Galactic latitude,
  showing GALEX FUV-NUV.
Models as described in the previous figure. 
In the right panels, composite model colors are added for WD+main sequence or WD+giant binaries, for WDs of {\mbox{$T_{\rm eff}$}~}=100,000K (purple)  and 30,000K (dark pink), and radii of R$_{WD}$=0.1{\mbox{$R_{\odot}$}}({log~{\it g}}=6) and  R$_{WD}$=0.02{\mbox{$R_{\odot}$}}({log~{\it g}}=8) with  less evolved companions of representative spectral types. The colors of a single WD with these parameters are also shown for reference. 
Also, QSO templates with enhanced {\mbox{Ly$_{\alpha}$}}
 The lower panels plot an optical color in the X-axis: note the different scale, and the much reduced sensitivity to discern  object types, and stellar gravities, with respect to UV-optical colors (top plots). The lower plots show a wider FUV-NUV range, to include cooler sources, the top plots have a narrower FUV-NUV range to better see the hot-WD binaries.   
\label{f_colcolsdss2} }
\end{figure*}

As mentioned, the match with SDSS offers, in addition to five optical bands extending the FUV, NUV GALEX measurements, also a morphological classification of the source shape (with respect to the PSF, thus limited to the SDSS $\sim$1.4{\mbox{$^{\prime\prime}$~}} resolution, which is higher than GALEX's). In Figures \ref{f_colcolsdss1} and  \ref{f_colcolsdss2} we color-coded the sources according to the SDSS classification of point-like sources ({\it type}=``STAR'') as blue dots, and extended sources ({\it type}=``GALAXY'') as black dots.  Note that SDSS {\it type}=''STAR'' refers to the source shape (sharpness), therefore it includes both stellar sources and extra-Galactic sources with a centrally peaked flux distribution, the latter being mostly QSOs. 

\begin{figure*}[!h]
\vskip -.72cm
\centerline{
\includegraphics[width=8.cm]{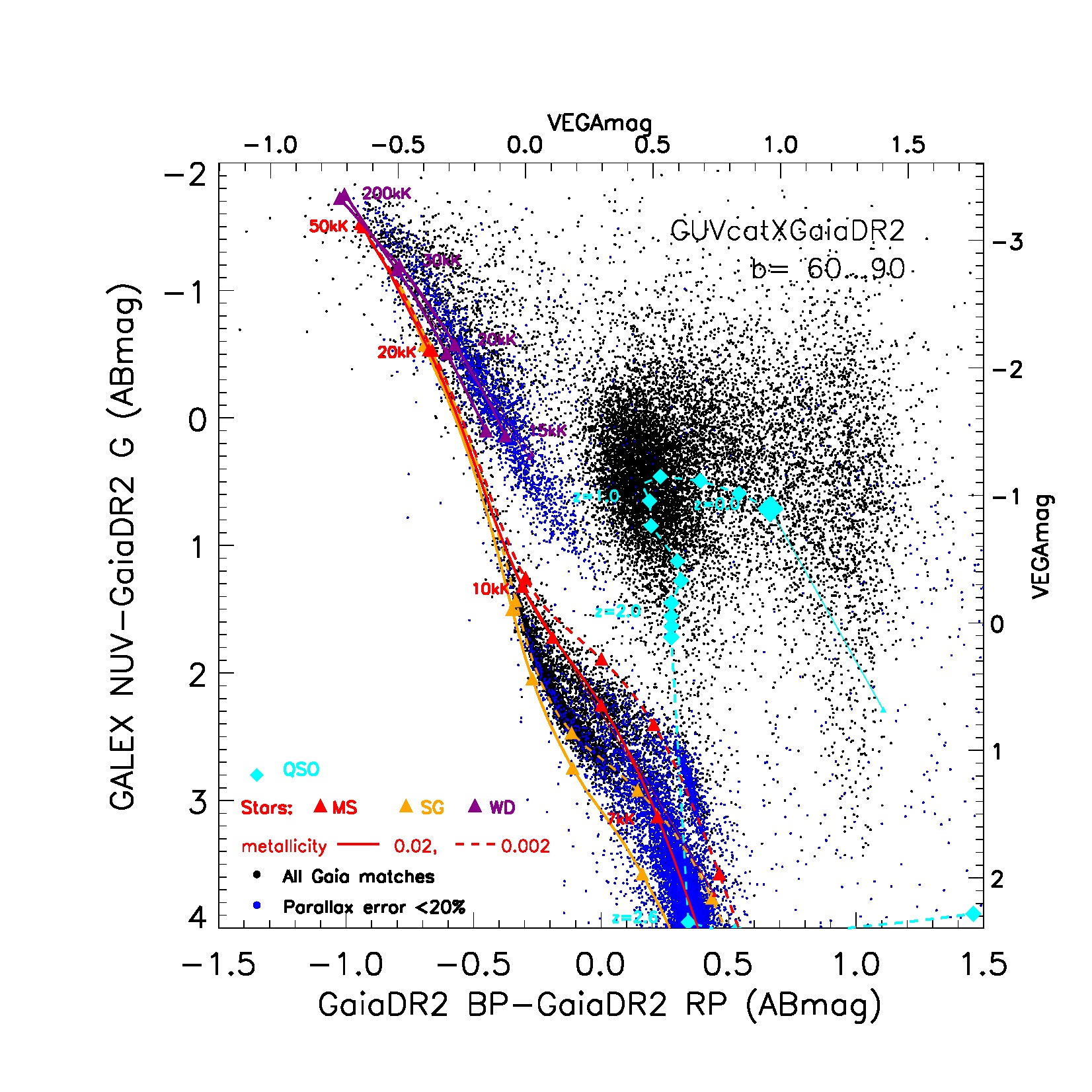} 
\includegraphics[width=8.cm]{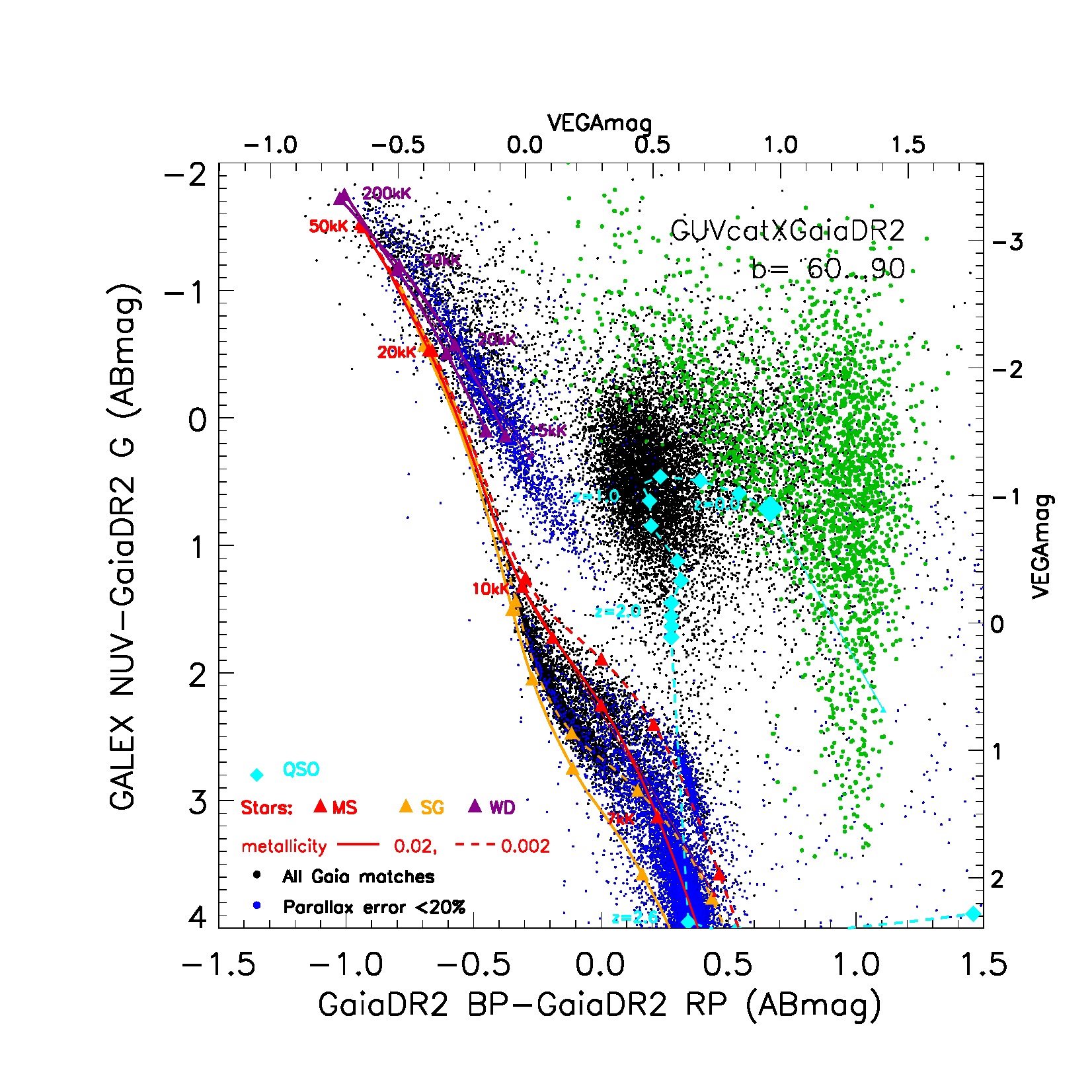}
}
\centerline{
\includegraphics[width=8.cm]{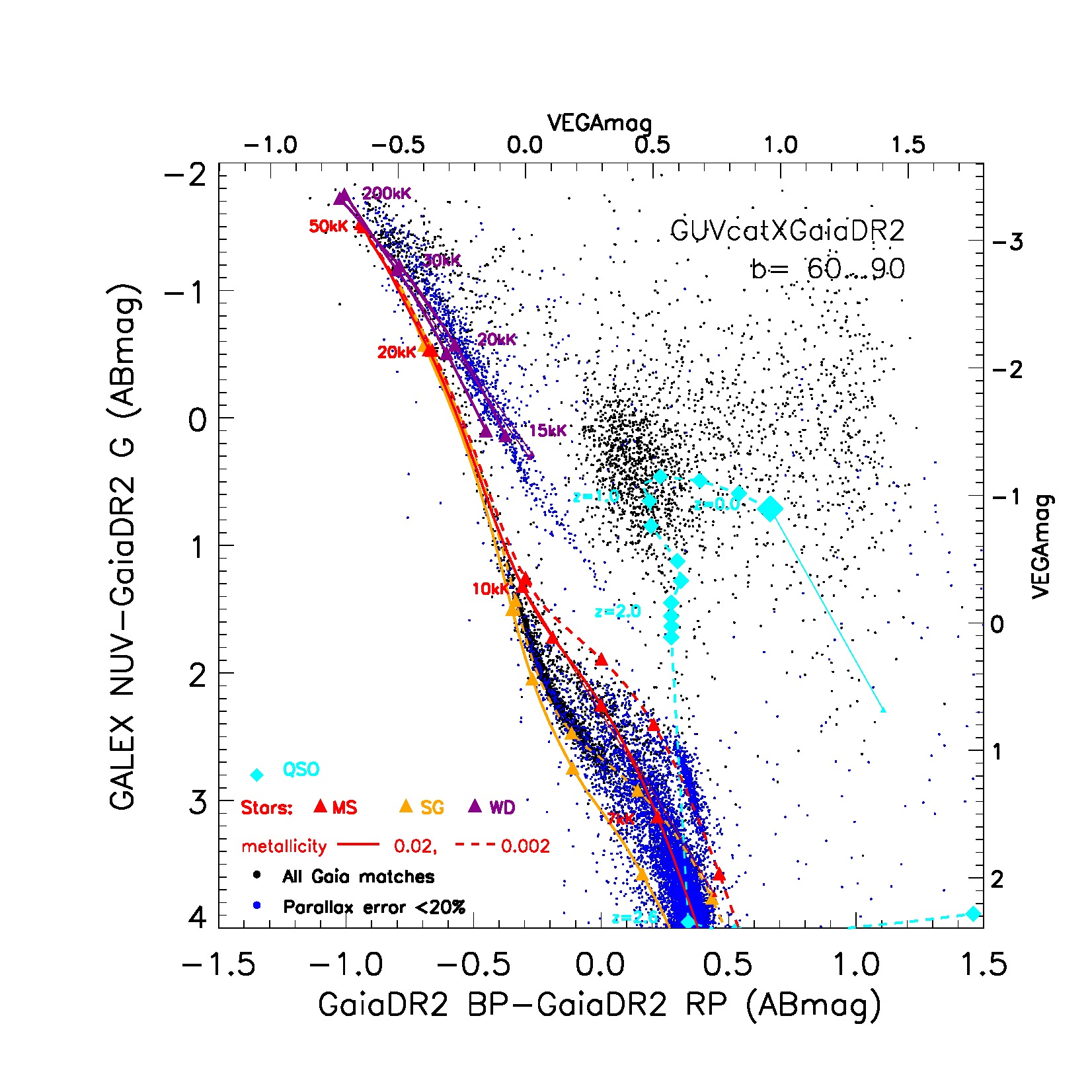}
\includegraphics[width=8.cm]{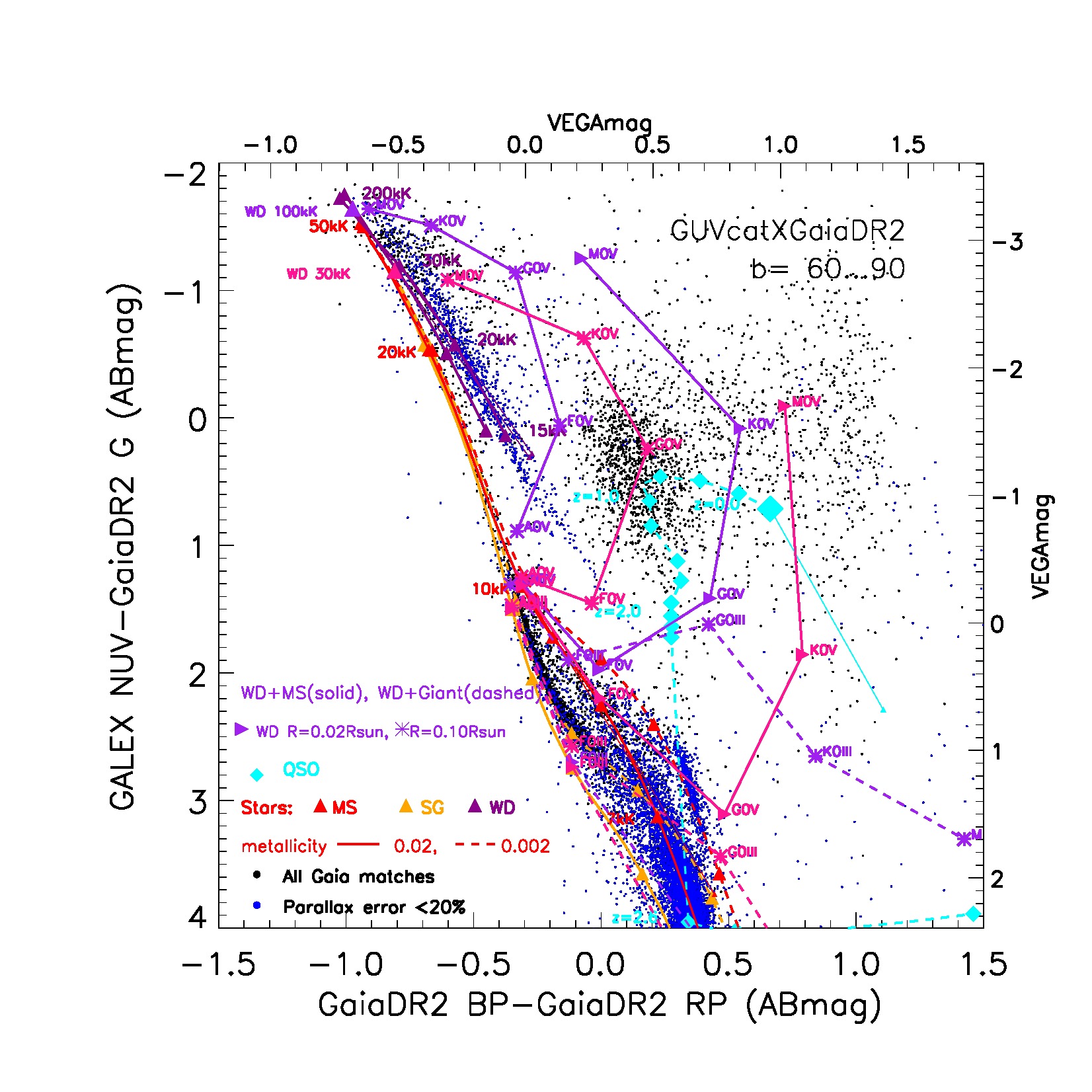} 
}
\vskip -0.3cm
\caption{Top row: color-color diagrams of $GUVmatch$\_AISxGaiaDR2 sources 
at b$\geq$60{\mbox{$^{\circ}$}}~N, with error $<$0.1,0.1,0.2,0.2mag in GALEX NUV, Gaia $G$, BP and RP mag.
Blue dots mark sources with parallax error $<$20\%. 
Green dots in the top-right panel mark ``extended'' sources (Section \ref{s_pntext}); they match the galaxies locus in previous plots (green dots are twice larger than others for evidence, as they are scant).  Second row, left: more stringent error cuts (0.05,0.05,0.1,0.1mag) 
reduce the spread around the loci defined by  model colors of object classes, as expected. It also reduces the relative number of extra-Galactic sources, that are typically fainter than Galactic objects, 
since progressively stringent error cuts translate into brighter magnitude limits (see e.g., \citet{bia07apj,bia11b}). The right-lower panel shows also composite colors for  WD+main sequence or WD+giant binaries for
 WD's {\mbox{$T_{\rm eff}$}~}=100,000K (purple) and 30,000K (dark pink),  R$_{WD}$=0.1{\mbox{$R_{\odot}$}} ({log~{\it g}}=6) and  R$_{WD}$=0.02{\mbox{$R_{\odot}$}} ({log~{\it g}}=8), with main-sequence and giant  companions of representative spectral types. 
\label{f_colcolgaia1}
}
\end{figure*}

\begin{figure*}[!h]
\centerline{
\includegraphics[width=9.cm]{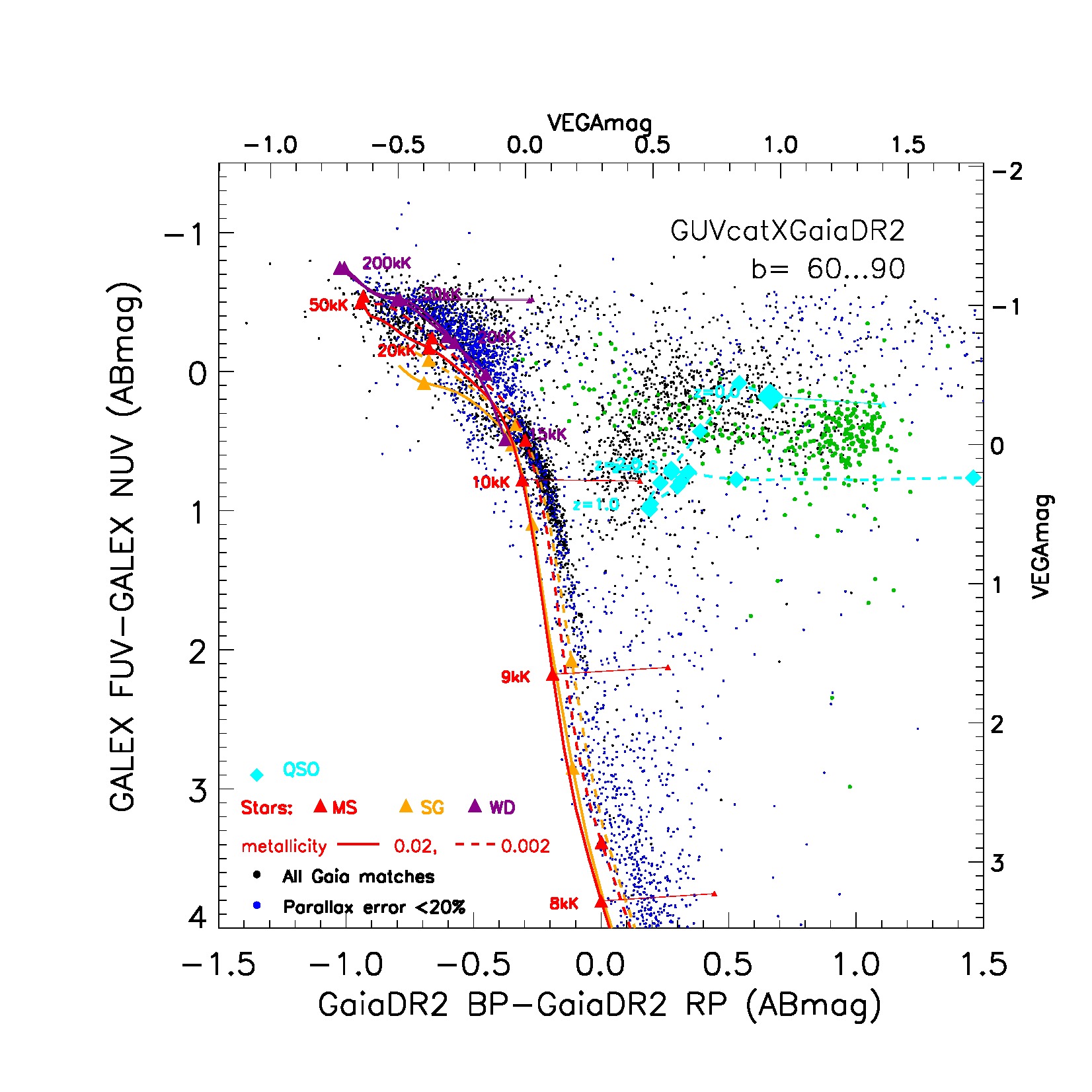}
\includegraphics[width=9.cm]{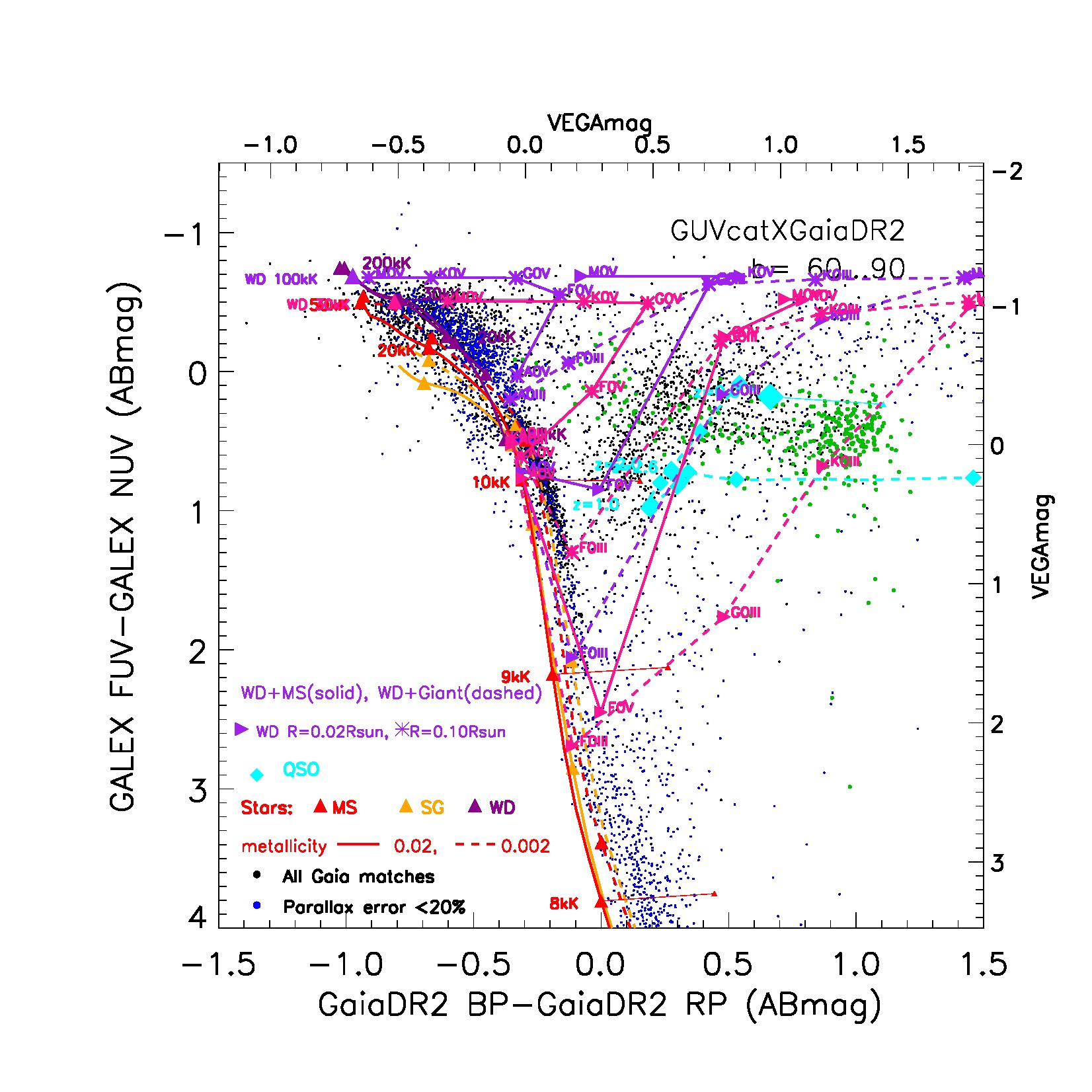} 
}
\vskip -0.3cm
\caption{Same data and symbols as in the previous figure. Plots show
the GALEX FUV-NUV color, that is essentially reddening-free for Milky~Way-type extinction with {$R_V$~}=3.1: note the reddening arrows, almost horizontal, while 
in the previous figure the reddening 
runs almost parallel to the stellar models' {\mbox{$T_{\rm eff}$}~} sequence, making it impossible to discern [Teff,{\mbox{$E_{B\!-\!V}$}}] effects, that are easily disentangled in this color combination.  Error cuts here are $<$0.1,0.1,0.1,0.2,0.2mag for GALEX FUV, NUV, Gaia-G, BP and RP respectively. About one tenth of the NUV sources are also detected in FUV. 
 As in previous figures, we show stellar and QSO model colors, with (right) and without (left) WD+cooler-star binaries for clarity. Blue points mark sources with parallax measurements (error $<$20\%), and green dots (left panel) mark extended sources (i.e. sources with a ``bad'' Gaia PSF fit), likely galaxies.  
\label{f_colcolgaia2}
}
\end{figure*}

\begin{figure*}[!h]
\centerline{
\includegraphics[width=9.cm]{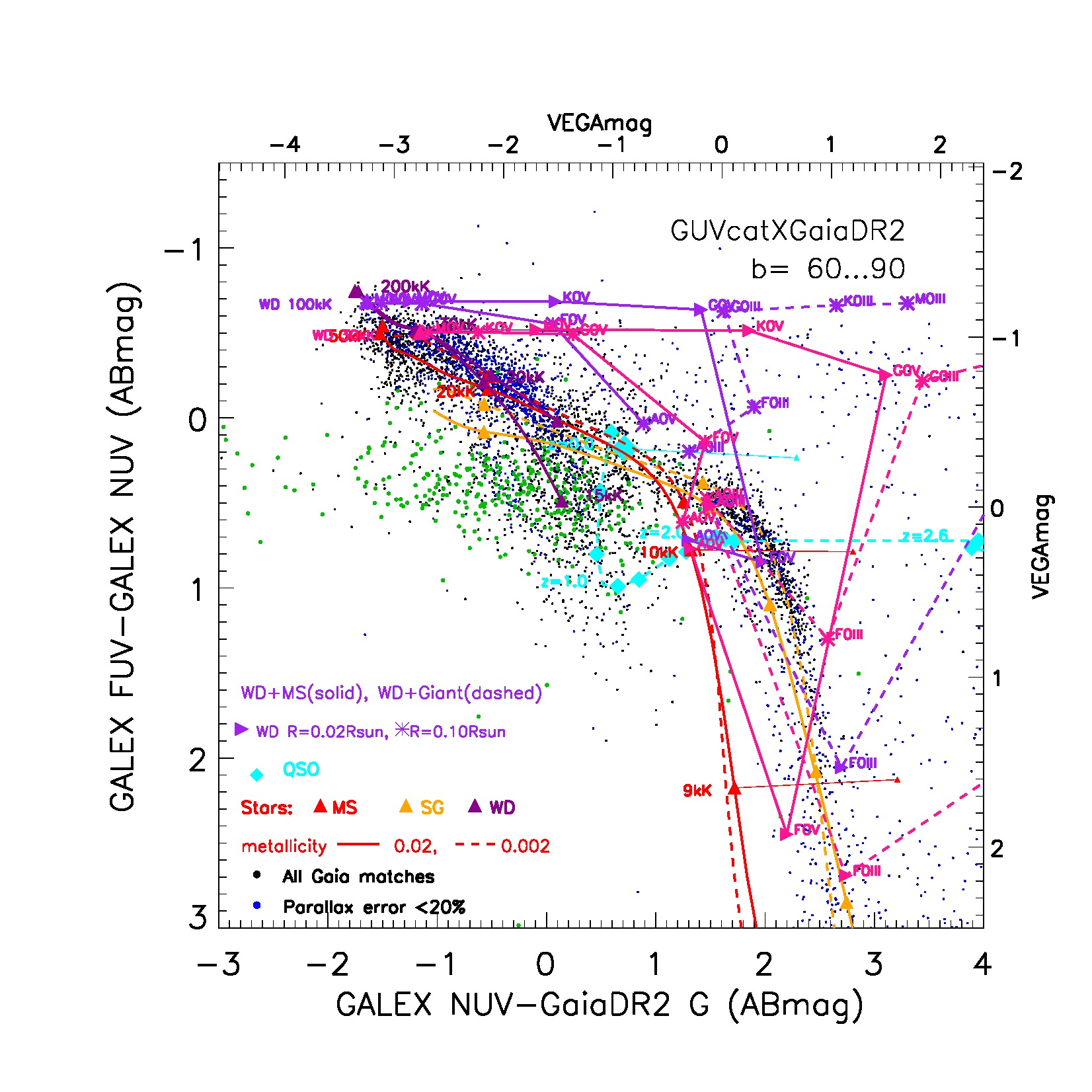}
\includegraphics[width=9.cm]{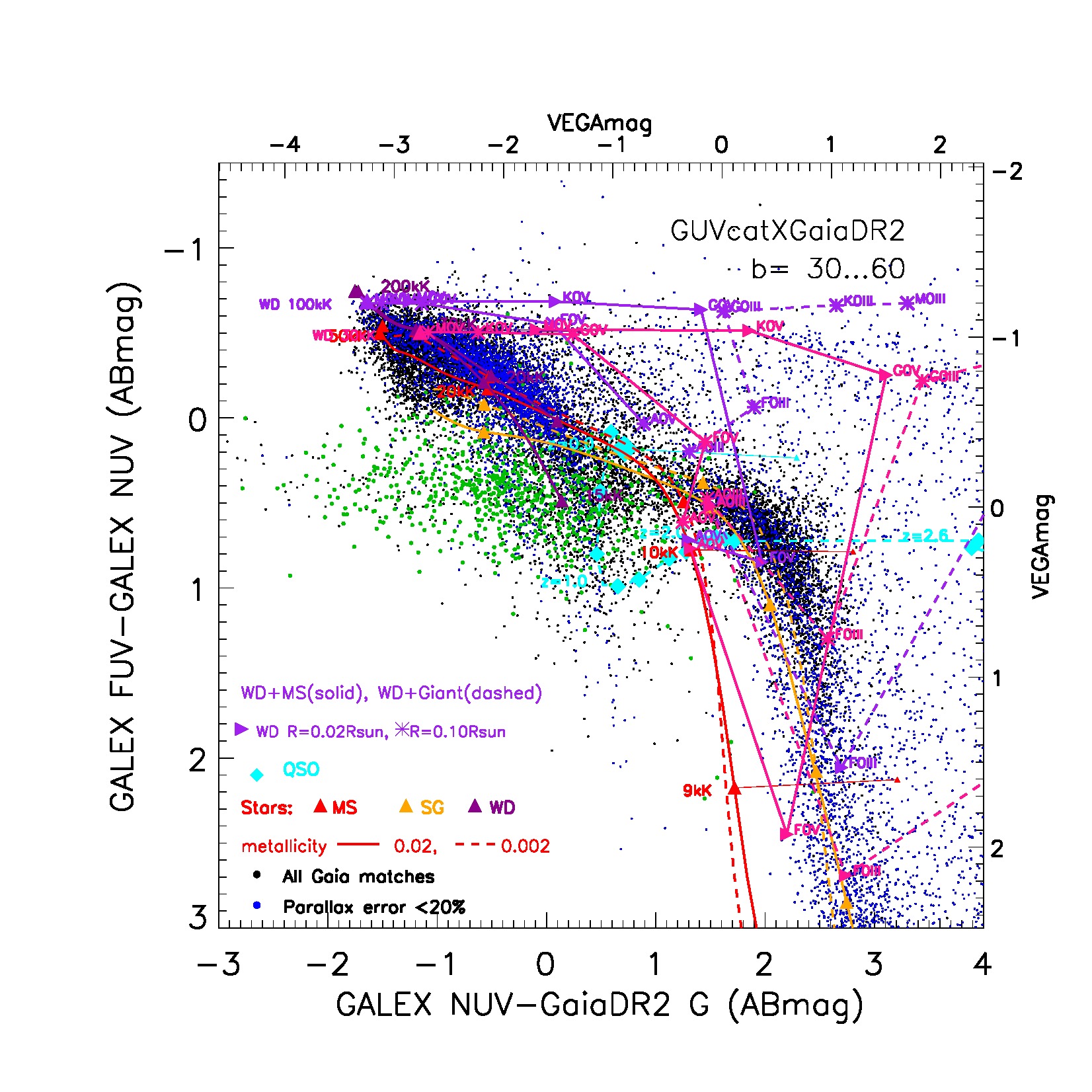}
}
\vskip -0.3cm
\caption{Data and symbols as in the previous figures, but this color combination separates red-shift$\sim$1 QSOs from binaries with a hot WD, while the single-star sequences cross each other in the high {\mbox{$T_{\rm eff}$}~} regime, and at cooler {\mbox{$T_{\rm eff}$}}'s they overlap with QSOs of red-shifts $\sim$0 and $\sim$2; the cooler ({\mbox{$T_{\rm eff}$}~}$\sim$15,000K) single WDs  are close to the QSOs color locus. The left panel shows the sample at Galatic latitudes b=90-60{\mbox{$^{\circ}$}} N, as in the $GUVmatch$\_AISxSDSSdr14 plots; the right panel shows the b=30-60{\mbox{$^{\circ}$}} N range.   Some of the sources extending to the left of the model sequences are probably galaxies, for which GALEX and Gaia photometry may be integrated over different areas, hence some probably meaningless very negative NUV - $G$ values.  Some may be emission-line nebulae (e.g.,  \citet{mgomez20}). 
\label{f_colcolgaia3}
}
\end{figure*}

  The match of the UV sources with Gaia adds measurements in up to three optical bands, Gaia's G, BP, and RP, and a parallax (and proper motion) measurement
 when a solution for these parameters was possible in Gaia DR2 (see Table \ref{t_statgaia} for the fraction of sources with parallax measurements).  In the color-color plots of the matched Gaia sources (Figures \ref{f_colcolgaia1} to \ref{f_colcolgaia3}), we  marked with blue dots the sources with existing parallax measurements that are not negative (these indicate a failed solution) and have parallax error better than 20\%. 
 
 The comparison of similar color combinations of UV-Gaia and UV-SDSS colors (Figures \ref{f_colcolsdss1} to  \ref{f_colcolgaia3}) shows similar distinct loci for stars, QSOs and galaxies, but marked differences in the relative source content. Most Gaia detections are point-like sources, and definitely more are stellar sources than QSOs; interestingly, Gaia also includes numerous cooler stars with good measurements, that are readily eliminated in the SDSS sample by error cuts. 
Conversely, SDSS samples show a much higher relative number of extended sources, that are likely galaxies, shown as black dots in Figures \ref{f_colcolsdss1} and \ref{f_colcolsdss2}.    Also interesting in the Gaia sample are the numerous stars cooler than $\sim$10,000K, showing a  well-defined sequence of likely sub-solar metallicity giants or super-giants (as in the SDSS plots), mostly lacking parallax measurements, as can be expected because supergiants can be detected out to large distances at high Galactic latitudes. The plots also show two well-populated stellar sequences at lower {\mbox{$T_{\rm eff}$}~}, that mostly do have good parallax measurements, indicative of their proximity 
that allows them to be detected in $GUVcat$\_AIS, in spite of their cool temperatures and low intrinsic UV luminosity.  As for the hot WDs, elusive at longer wavelengths but easily identified with the UV catalog, the two magenta sequences of WD model colors, with {log~{\it g}}~=7 and 9 and {\mbox{$T_{\rm eff}$}~} between 200,000K and 15,000K, encompass a clear, well-defined sequence of Milky Way hot-WD candidates, that \citet{bia11a} found to have a purity of over 99\%,  by checking a subsample of about 4000 such  sources with SDSS spectra.  

The stellar models colors are from the Kurucz and TLUSTY model grids described by \citet{bia09,bia11,bia07apj,bia11a}. The QSO model colors, including colors computed from templates with {\mbox{Ly$_{\alpha}$}}

\subsection{Stellar binaries.}
\label{s_bina}

 In some color-color plots we also show the loci of binaries composed of a hot WD and a cooler stellar companion. Examples are shown for WDs of {\mbox{$T_{\rm eff}$}~}=100,000K and 30,000K (purple and dark-pink models, respectively); for each {\mbox{$T_{\rm eff}$}~},  two WD radii, R$_{WD}$=0.1{\mbox{$R_{\odot}$}} (with {log~{\it g}}=6) and  R$_{WD}$=0.02{\mbox{$R_{\odot}$}} (with {log~{\it g}}=8), are used to compute the binary composite magnitudes.  
  {\mbox{$T_{\rm eff}$}~} and Radii for the cool-star companions (main sequence or giants) of representative spectral types  are taken from standard compilations. Given the scale of the plots, that span several magnitudes, using sligtly different calibrations for spectral types will not produce an appreciable difference.  When optical colors are involved, the addition of a cooler star displaces the source towards redder colors with respect to a single hot WD, as expected, the extent of this effect depends on the ratio of radii and {\mbox{$T_{\rm eff}$}~} between the stellar pair. Instead, the  FUV-NUV color of the hot~WD is hardly affected by cooler companions, again depending on how cool and how large the companion is.  Such effects, initially pointed out by our earlier work, and used by \citet{bia18b} to select binaries for HST follow-up observations, enable selection of otherwise elusive binaries with a hot (and optically faint) WD.  
While the single-star sequences have a smooth, almost linear color progression with {\mbox{$T_{\rm eff}$}~}, the colors of binaries  show a complex behaviour as a function of the cool companion's {\mbox{$T_{\rm eff}$}~}, owing to the differing relative contribution of the two components in each filter. This can be understood, for example,  by looking at examples of binary SEDs shown by  \citet{bia18b} and Bianchi et al. (in preparation).

\section{Summary and Conclusions}
\label{s_summary}

We have matched the 83~million UV sources of $GUVcat$\_AIS\_FOV055 \citep{bia17guvcat,bia20guvcat} with the optical databases  Gaia DR2 and SDSS DR14.
With a match radius of 3{\mbox{$^{\prime\prime}$~}}, we obtained:  
\begin{itemize}
\item
{\bf $GUVmatch$\_AISxGaiaDR2}. 
30,024,791 
$GUVcat$\_AIS UV sources have 31,925,294 
Gaia DR2 counterparts over the entire $GUVcat$\_AIS footprint, 
  26,275,572 
of them have a parallax measurement, 18,588,139 with a parallax error  $\leq$30\%.  FUV and  NUV  magnitudes are given in AB~mag, and Gaia's mags G, BP, RP in the Vega-mag system, to keep the values unaltered from the original respective databases. Conversions between AB and Vega mag systems are given in Table \ref{t_vegaab}.  

\item
{\bf $GUVmatch$\_AISxSDSSdr14}. 
 22,207,563 
$GUVcat$\_AIS sources have 
23,310,532 SDSS DR14 counterparts, 10,167,460 
of which are point-like, over a total overlapping sky area of $\sim$11,100~square degrees.  All magnitudes FUV, NUV, {\it u, g, r, i, z} are given in the AB mag system.

\end{itemize}

The matched catalogs include all columns (tags)  from  $GUVcat$\_AIS, described in Table 8 of \citet{bia17guvcat},  and all tags from the  optical databases, described in their respective web sites, as well as additional tags to track multiple matches,  defined in Table \ref{t_mmtags}.   Two $GUVcat$  tags,  $INLARGEOBJ$ and $LARGEOBJSIZE$, flag sources that fall within the footprint of extended, complex objects, where the photometry of individual point sources may be compromised, or the flux may happen to be measured over different areas in different filters, making the colors misleading (see \citet{bia17guvcat}). 
~\\

\noindent
{\it Artifacts}

 We did not remove $GUVcat$ sources with artifacts nor SDSS measurements with saturation flags or other warnings, because recipes to clean artifacts should be optimized according to each specific analysis and objective.  Sources are included regardless of  artifact flags, such as saturation or instrumental hot pixels, etc., also because an artifact may affect one filter but not the others.  Some discussion about artifacts in GALEX photometry is given by \citet{bia17guvcat}. The number of affected sources, however,  is very small, therefore irrelevant for the purpose of our overview.

\noindent
{\it Multiple Matches} 

  A fraction of the $GUVcat$ sources have more than one optical counterpart, on average $\sim$6\% for Gaia DR2, $\sim$5\% for SDSS DR14; 
this fraction can highly vary with magnitude range and Galactic latitude (see \citet{bia11a,bia11b}, and Tables \ref{t_statgaia} and  \ref{t_statsdss}).  
All counterparts within  the match radius are included in the catalogs, to enable tracking of multiple matches (Table \ref{t_mmtags}). 
This is important for SED analysis, since a GALEX source that is resolved in two or more optical counterparts may include the composite flux of all of these counterparts, while the optical magnitude of the closest optical match will only include the flux from the one closest match, making UV$-$optical colors biased.  
 
\noindent
{\it Sky coverage. Source density}
 
To estimate, from source counts of an extracted sample, the average surface density of sources in the matched databases presented here,  or that of $GUVcat$, SDSS, and Gaia,  
 the areas of overlap between the footprint of the matched catalogs can be calculated for any region of the sky with the online interactive tool $AREAcat$\footnote{\url{http://dolomiti.pha.jhu.edu/uvsky/area/AREAcat.php}} described by \citet{bia19areacat}  who also give area coverage at different Galactic latitudes for the matched catalogs, as well as for the individual databases, in their Table 1.

\noindent
{\it Sources in the footprint of extended objects} 

 The number of matched sources, of subsets with single matches, as well as of those with parallax measurements for Gaia DR2, or with point-like classification and with existing spectra for SDSS, 
are given in Tables \ref{t_statgaia} and \ref{t_statsdss}, 
by Galactic latitude and in total. In these tables we list  counts both including all sources, and excluding those that fall within extended objects (large galaxies or clusters) larger than 30~arcmin, because sources in such complicated regions may have problematic identification and measurements (see \citet{bia17guvcat}, in particular their Figure 5).  
Also, the flux of such sources maybe have been measured by GALEX, Gaia, and SDSS by integrating over very different  profiles, making colors from combined datasets inconsistent and meaningless. Such sources require custom-made photometry, as shown by \citet{bia17guvcat}.

\subsection{How to access the Matched Catalogs and Related Tools}
\label{s_access} 

The matched catalogs can be downloaded (as fits or csv files) from the author web site \url{http://dolomiti.pha.jhu.edu/uvsky}, where $GUVcat$\_AIS is also available, as well as  the AREAcat tool to calculate sky coverage of each matched catalog (and of the individual databases) in any chosen region of the sky.  Both   $GUVcat$ and the cross-matched catalogs are also  available at MAST as a High Level Science Product at \dataset[https://doi.org/10.17909/t9-pyxy-kg53]{https://doi.org/10.17909/t9-pyxy-kg53}. 

 In addition, the  catalogs  can be conveniently browsed through the Casjobs interface at MAST (\url{http://mastweb.stsci.edu/gcasjobs/}), and will also be available from the SIMBAD Astronomical Database (CDS, Strasbourg) through Vizier;  these sites also enable direct cross matching with additional catalogs residing in their respective repositories or uploaded by the user.

{}
  
\acknowledgments 
We acknowledge support from  NASA grants ADAP 80NSSC19K0527 and NNX17AF35G.  We are grateful to Chase Million and David Thilker whom we often consult regarding GALEX data issues, for their always  helpful answers.  We thank especially Sihao Cheng for sharing results of his tests for Gaia pointlike/extended source classification, and Carla Cacciari and Anthony Brown for  useful  Gaia suggestions. We thank the referee for a careful and timely reading of the manuscript and helpful suggestions.  

{\it Facilities:} \facility{GALEX},  \facility{MAST} 

\textbf{ORCID iDs}
Luciana Bianchi https://orcid.org/0000-0001-7746-5461

\newpage

\begin{deluxetable}{lrrrrrrrrr}
\tabletypesize{\tiny}
\tablecaption{Match $GUVcat\_AIS$ x Gaia DR2: Statistics.\tablenotemark{a} \label{t_statgaia} }
\tablehead{
\colhead{$b$ range} & \multicolumn{3}{c}{------- \# matches -------} &  \colhead{Multiple } & \colhead{Reverse}
 & \colhead{\# with} & \multicolumn{3}{c}{\# parallax error $\leq$} \\
 & \multicolumn{1}{c}{total} & \multicolumn{1}{c}{unique UV} & \colhead{single} & \colhead{matches} & \colhead{mult.mtc} 
 & \colhead{parallax} & \colhead{20\%}   & \colhead{30\%} & \colhead{50\%} \\
 &   \colhead{matches}               & \colhead{sources}  & \colhead{matches}&  \colhead{\# (\%)} & \colhead{\# (\%)} 
 & \colhead{$>$ 0.}   & \colhead{ }      & \colhead{} & \colhead{} 
}
\tablewidth{0pt}
\startdata
90...85 & 
37166 & 36505 & 35854 & 651 (01.78\%)  & 231 (00.63\%) &
28052 & 14913 & 16801 & 19433 \\ 
 & 
35904 & 35258 & 34622 & 1883 (01.80\%)  & 226 (00.64\%) & 
27110 & 14428 & 16247 & 18785 \\ 
85...80 & 
100390 & 98528 & 96683 & 1845 (01.87\%)  & 614 (00.62\%) &
77305 & 43102 & 48179 & 55250 \\ 
 & 
96313 & 94504 & 92711 & 5817 (01.90\%)  & 592 (00.63\%) & 
74211 & 41308 & 46212 & 53033 \\ 
80...75 & 
179356 & 175393 & 171537 & 3856 (02.20\%)  & 1103 (00.63\%) &
135580 & 74739 & 83665 & 95742 \\ 
 & 
177431 & 173856 & 170322 & 5071 (02.03\%)  & 1050 (00.60\%) & 
134661 & 74695 & 83612 & 95611 \\ 
75...70 & 
257933 & 252914 & 247945 & 4969 (01.96\%)  & 1641 (00.65\%) &
195890 & 109940 & 123036 & 140652 \\ 
 & 
257933 & 252914 & 247945 & 4969 (01.96\%)  & 1641 (00.65\%) & 
195890 & 109940 & 123036 & 140652 \\ 
70...65 & 
315831 & 309644 & 303500 & 6144 (01.98\%)  & 2128 (00.69\%) &
243584 & 137750 & 154507 & 176575 \\ 
 & 
315831 & 309644 & 303500 & 6144 (01.98\%)  & 2128 (00.69\%) & 
243584 & 137750 & 154507 & 176575 \\ 
65...60 & 
399685 & 391622 & 383621 & 8001 (02.04\%)  & 2454 (00.63\%) &
311052 & 174758 & 196835 & 225933 \\ 
 & 
399685 & 391622 & 383621 & 8001 (02.04\%)  & 2454 (00.63\%) & 
311052 & 174758 & 196835 & 225933 \\ 
60...55 & 
522010 & 511189 & 500449 & 10740 (02.10\%)  & 3250 (00.64\%) &
411617 & 234077 & 263893 & 302527 \\ 
 & 
522010 & 511189 & 500449 & 10740 (02.10\%)  & 3250 (00.64\%) & 
411617 & 234077 & 263893 & 302527 \\ 
55...50 & 
624112 & 610401 & 596798 & 13603 (02.23\%)  & 4015 (00.66\%) &
502596 & 294695 & 332267 & 378942 \\ 
 & 
624112 & 610401 & 596798 & 13603 (02.23\%)  & 4015 (00.66\%) & 
502596 & 294695 & 332267 & 378942 \\ 
50...45 & 
800004 & 780943 & 762090 & 18853 (02.41\%)  & 4913 (00.63\%) &
648602 & 381248 & 430671 & 492663 \\ 
 & 
799075 & 780198 & 761509 & 19434 (02.40\%)  & 4912 (00.63\%) & 
648052 & 381206 & 430600 & 492506 \\ 
45...40 & 
1014713 & 988382 & 962346 & 26036 (02.63\%)  & 6844 (00.69\%) &
833513 & 497627 & 564076 & 644634 \\ 
 & 
1014713 & 988382 & 962346 & 26036 (02.63\%)  & 6844 (00.69\%) & 
833513 & 497627 & 564076 & 644634 \\ 
40...35 & 
1225590 & 1189097 & 1153071 & 36026 (03.03\%)  & 8200 (00.69\%) &
1030336 & 629021 & 712460 & 811868 \\ 
 & 
1225564 & 1189072 & 1153047 & 36050 (03.03\%)  & 8200 (00.69\%) & 
1030313 & 629003 & 712442 & 811848 \\ 
35...30 & 
1520896 & 1468522 & 1416965 & 51557 (03.51\%)  & 9809 (00.67\%) &
1296973 & 803431 & 911550 & 1036984 \\ 
 & 
1514477 & 1462291 & 1410921 & 57601 (03.51\%)  & 9777 (00.67\%) & 
1291457 & 799505 & 907277 & 1032306 \\ 
30...25 & 
1776297 & 1702747 & 1630787 & 71960 (04.23\%)  & 10741 (00.63\%) &
1529460 & 975174 & 1104504 & 1248778 \\ 
 & 
1776297 & 1702747 & 1630787 & 71960 (04.23\%)  & 10741 (00.63\%) & 
1529460 & 975174 & 1104504 & 1248778 \\ 
25...20 & 
2039885 & 1937607 & 1838059 & 99548 (05.14\%)  & 12302 (00.63\%) &
1764979 & 1159053 & 1311694 & 1473283 \\ 
 & 
2039010 & 1936791 & 1837297 & 100310 (05.14\%)  & 12294 (00.63\%) & 
1764192 & 1158378 & 1310989 & 1472546 \\ 
20...15 & 
2158395 & 2017830 & 1882301 & 135529 (06.72\%)  & 11844 (00.59\%) &
1862541 & 1243181 & 1419498 & 1590218 \\ 
 & 
2150952 & 2011003 & 1876044 & 141786 (06.71\%)  & 11798 (00.59\%) & 
1856173 & 1238770 & 1414505 & 1584661 \\ 
15...10 & 
2096516 & 1887520 & 1693878 & 193642 (10.26\%)  & 9503 (00.50\%) &
1751890 & 1171372 & 1350013 & 1516601 \\ 
 & 
2094443 & 1885668 & 1692235 & 195285 (10.26\%)  & 9486 (00.50\%) & 
1750186 & 1170159 & 1348664 & 1515128 \\ 
10...5 & 
882090 & 755647 & 646590 & 109057 (14.43\%)  & 3366 (00.45\%) &
695727 & 463666 & 534592 & 602463 \\ 
 & 
882090 & 755647 & 646590 & 109057 (14.43\%)  & 3366 (00.45\%) & 
695727 & 463666 & 534592 & 602463 \\ 
5...0 & 
87028 & 75340 & 64843 & 10497 (13.93\%)  & 279 (00.37\%) &
69380 & 49251 & 55684 & 61395 \\ 
 & 
86328 & 74733 & 64326 & 11014 (13.93\%)  & 267 (00.36\%) & 
68822 & 48885 & 55256 & 60915 \\ 
0...-5 & 
49725 & 45168 & 40808 & 4360 (09.65\%)  & 118 (00.26\%) &
41924 & 32626 & 35483 & 38096 \\ 
 & 
49609 & 45068 & 40722 & 4446 (09.64\%)  & 118 (00.26\%) & 
41829 & 32546 & 35399 & 38009 \\ 
-5...-10 & 
441984 & 361938 & 293239 & 68699 (18.98\%)  & 1620 (00.45\%) &
330249 & 245362 & 270611 & 293339 \\ 
 & 
440855 & 360908 & 292303 & 69635 (19.01\%)  & 1619 (00.45\%) & 
329272 & 244530 & 269727 & 292420 \\ 
-10...-15 & 
1584118 & 1363316 & 1164738 & 198578 (14.57\%)  & 6422 (00.47\%) &
1254651 & 862572 & 970059 & 1077024 \\ 
 & 
1518247 & 1318639 & 1136577 & 226739 (13.81\%)  & 6273 (00.48\%) & 
1218184 & 846583 & 950736 & 1052981 \\ 
-15...-20 & 
2240184 & 2013886 & 1801929 & 211957 (10.52\%)  & 10994 (00.55\%) &
1827972 & 1112037 & 1280319 & 1471552 \\ 
 & 
2147708 & 1938593 & 1741839 & 272047 (10.15\%)  & 10606 (00.55\%) & 
1763584 & 1081941 & 1244007 & 1426829 \\ 
-20...-25 & 
2022316 & 1894665 & 1771469 & 123196 (06.50\%)  & 12572 (00.66\%) &
1722343 & 1062453 & 1218297 & 1394834 \\ 
 & 
2017783 & 1890402 & 1767458 & 127207 (06.50\%)  & 12555 (00.66\%) & 
1718829 & 1060188 & 1215735 & 1391943 \\ 
-25...-30 & 
1901493 & 1799945 & 1702177 & 97768 (05.43\%)  & 13174 (00.73\%) &
1588977 & 947566 & 1087137 & 1250466 \\ 
 & 
1821160 & 1733444 & 1647945 & 152000 (04.93\%)  & 12819 (00.74\%) & 
1551262 & 936073 & 1074369 & 1235072 \\ 
-30...-35 & 
1792601 & 1671312 & 1561854 & 109458 (06.55\%)  & 12097 (00.72\%) &
1409764 & 810207 & 927073 & 1067346 \\ 
 & 
1554521 & 1492890 & 1432521 & 238791 (04.04\%)  & 10880 (00.73\%) & 
1316529 & 790952 & 905488 & 1039274 \\ 
-35...-40 & 
1304958 & 1256064 & 1208755 & 47309 (03.77\%)  & 10273 (00.82\%) &
1080221 & 652166 & 740868 & 845429 \\ 
 & 
1246228 & 1205727 & 1165906 & 90158 (03.30\%)  & 10116 (00.84\%) & 
1051101 & 642369 & 730139 & 832730 \\ 
-40...-45 & 
1085720 & 1046294 & 1008631 & 37663 (03.60\%)  & 8031 (00.77\%) &
871944 & 514106 & 583229 & 666170 \\ 
 & 
1039739 & 1009051 & 978866 & 67428 (02.99\%)  & 7893 (00.78\%) & 
852042 & 511107 & 579242 & 659852 \\ 
-45...-50 & 
840241 & 816361 & 793077 & 23284 (02.85\%)  & 7465 (00.91\%) &
679217 & 405908 & 458103 & 521787 \\ 
 & 
819236 & 798057 & 777129 & 39232 (02.62\%)  & 7381 (00.92\%) & 
668387 & 403591 & 455143 & 517496 \\ 
-50...-55 & 
659448 & 643777 & 628237 & 15540 (02.41\%)  & 6489 (01.01\%) &
537432 & 327411 & 367814 & 416506 \\ 
 & 
657328 & 641713 & 626229 & 17548 (02.41\%)  & 6472 (01.01\%) & 
535709 & 326219 & 366532 & 415109 \\ 
-55...-60 & 
569548 & 556881 & 544316 & 12565 (02.26\%)  & 4720 (00.85\%) &
458046 & 277814 & 311183 & 351388 \\ 
 & 
564344 & 551783 & 539323 & 17558 (02.26\%)  & 4692 (00.85\%) & 
453889 & 275039 & 308165 & 348101 \\ 
-60...-65 & 
426160 & 417042 & 407988 & 9054 (02.17\%)  & 3120 (00.75\%) &
335976 & 201877 & 225698 & 254950 \\ 
 & 
426160 & 417042 & 407988 & 9054 (02.17\%)  & 3120 (00.75\%) & 
335976 & 201877 & 225698 & 254950 \\ 
-65...-70 & 
365868 & 358039 & 350339 & 7700 (02.15\%)  & 3082 (00.86\%) &
286011 & 172493 & 191825 & 216321 \\ 
 & 
365868 & 358039 & 350339 & 7700 (02.15\%)  & 3082 (00.86\%) & 
286011 & 172493 & 191825 & 216321 \\ 
-70...-75 & 
268228 & 262655 & 257114 & 5541 (02.11\%)  & 1898 (00.72\%) &
206818 & 123373 & 137482 & 155303 \\ 
 & 
268228 & 262655 & 257114 & 5541 (02.11\%)  & 1898 (00.72\%) & 
206818 & 123373 & 137482 & 155303 \\ 
-75...-80 & 
182037 & 178105 & 174204 & 3901 (02.19\%)  & 1853 (01.04\%) &
139163 & 83798 & 92896 & 104394 \\ 
 & 
181590 & 177666 & 173773 & 4332 (02.19\%)  & 1853 (01.04\%) & 
138803 & 83590 & 92655 & 104123 \\ 
-80...-85 & 
117051 & 114601 & 112172 & 2429 (02.12\%)  & 1291 (01.13\%) &
88674 & 52683 & 58365 & 65759 \\ 
 & 
116753 & 114307 & 111882 & 2719 (02.12\%)  & 1291 (01.13\%) & 
88421 & 52527 & 58190 & 65562 \\ 
-85...-90 & 
35717 & 34911 & 34123 & 788 (02.26\%)  & 287 (00.82\%) &
27113 & 16055 & 17772 & 20024 \\ 
 & 
35717 & 34911 & 34123 & 788 (02.26\%)  & 287 (00.82\%) & 
27113 & 16055 & 17772 & 20024 \\ 
90...-90 & 
31925294 & 30024791 & 28242487 & 1782304 (05.94\%)  & 198743 (00.66\%) &
26275572 & 16357505 & 18588139 & 21084629 \\ 
 & 
31283242 & 29516815 & 27847107 & 2177684 (05.66\%)  & 195996 (00.66\%) & 
25952375 & 16245077 & 18457818 & 20923942 \\ 
\enddata
\tablenotetext{a}{For each Galactic latitude range, the second row gives counts excluding sources in extended objects such as galaxies or stellar clusters larger than 30arcmin (see \citet{bia17guvcat}). 
Counts of sources with parallax measurements refer to the unique GUVcat sources ($DISTANCERANK$$<$2), not counting multiple matches to the same sources.
    }
\end{deluxetable}

\newpage
\begin{deluxetable}{lrrrrrrrrrr}
\tabletypesize{\tiny}
\tablecaption{Match $GUVcat\_AIS$ x SDSS DR14: Statistics.\tablenotemark{a} \label{t_statsdss} }
\tablehead{
\colhead{$b$ range} & \multicolumn{3}{c}{------- \# matches -------} &  \colhead{Multiple } & \colhead{Reverse}
 & \multicolumn{2}{c}{\# Point-like}  & \multicolumn{3}{c}{\# SDSS spectra} \\
 & \multicolumn{1}{c}{total} & \multicolumn{1}{c}{unique UV} & \colhead{single} & \colhead{matches} & \colhead{mult.mtc} 
 & \colhead{ unique } & \colhead{ FUV}   & \colhead{ unique } & \colhead{ FUV} & \colhead{pnt} \\
 &   \colhead{matches}               & \colhead{sources}  & \colhead{matches}&  \colhead{\# (\%)} & \colhead{\# (\%)} 
 &                    &                  & \colhead{} & \colhead{} & 
}
\tablewidth{0pt}
\startdata
90...85 & 
172873 & 164978 & 157341 & 7637 (04.63\%)  & 1462 (00.89\%) &
41007 & 3261 & 7087 & 3441 & 2882 \\ 
 & 
167289 & 159632 & 152228 & 12750 (04.64\%)  & 1437 (00.90\%) & 
39610 & 3159 & 6882 & 3441 & 2812 \\ 
85...80 & 
397856 & 379062 & 360895 & 18167 (04.79\%)  & 3457 (00.91\%) &
106514 & 8784 & 19467 & 10603 & 7720 \\ 
 & 
383207 & 365182 & 347758 & 31304 (04.77\%)  & 3355 (00.92\%) & 
102416 & 8440 & 18638 & 10603 & 7490 \\ 
80...75 & 
722113 & 687621 & 654348 & 33273 (04.84\%)  & 5995 (00.87\%) &
190572 & 13285 & 35623 & 17533 & 14973 \\ 
 & 
721333 & 686895 & 653675 & 33946 (04.84\%)  & 5965 (00.87\%) & 
189951 & 13267 & 35523 & 17533 & 14877 \\ 
75...70 & 
1004931 & 959561 & 915741 & 43820 (04.57\%)  & 8841 (00.92\%) &
276126 & 20058 & 52478 & 26814 & 22026 \\ 
 & 
1004931 & 959561 & 915741 & 43820 (04.57\%)  & 8841 (00.92\%) & 
276126 & 20058 & 52478 & 26814 & 22026 \\ 
70...65 & 
1165379 & 1112613 & 1061650 & 50963 (04.58\%)  & 10366 (00.93\%) &
334967 & 24583 & 64906 & 31849 & 29336 \\ 
 & 
1165379 & 1112613 & 1061650 & 50963 (04.58\%)  & 10366 (00.93\%) & 
334967 & 24583 & 64906 & 31849 & 29336 \\ 
65...60 & 
1435902 & 1370927 & 1308205 & 62722 (04.58\%)  & 12467 (00.91\%) &
424597 & 30268 & 75321 & 37077 & 34068 \\ 
 & 
1435902 & 1370927 & 1308205 & 62722 (04.58\%)  & 12467 (00.91\%) & 
424597 & 30268 & 75321 & 37077 & 34068 \\ 
60...55 & 
1618031 & 1548624 & 1481453 & 67171 (04.34\%)  & 13564 (00.88\%) &
494329 & 33595 & 79899 & 40134 & 34414 \\ 
 & 
1618031 & 1548624 & 1481453 & 67171 (04.34\%)  & 13564 (00.88\%) & 
494329 & 33595 & 79899 & 40134 & 34414 \\ 
55...50 & 
1479736 & 1416075 & 1354386 & 61689 (04.36\%)  & 13259 (00.94\%) &
511309 & 32796 & 72104 & 35390 & 33506 \\ 
 & 
1479736 & 1416075 & 1354386 & 61689 (04.36\%)  & 13259 (00.94\%) & 
511309 & 32796 & 72104 & 35390 & 33506 \\ 
50...45 & 
1601683 & 1530855 & 1462398 & 68457 (04.47\%)  & 13769 (00.90\%) &
573799 & 35171 & 69282 & 33333 & 31588 \\ 
 & 
1601182 & 1530410 & 1462006 & 68849 (04.47\%)  & 13768 (00.90\%) & 
573412 & 35153 & 69273 & 33333 & 31584 \\ 
45...40 & 
1592536 & 1521609 & 1453149 & 68460 (04.50\%)  & 14281 (00.94\%) &
640150 & 34762 & 61967 & 28776 & 29099 \\ 
 & 
1592536 & 1521609 & 1453149 & 68460 (04.50\%)  & 14281 (00.94\%) & 
640150 & 34762 & 61967 & 28776 & 29099 \\ 
40...35 & 
1389482 & 1325368 & 1263605 & 61763 (04.66\%)  & 11499 (00.87\%) &
647806 & 34199 & 54402 & 24959 & 26887 \\ 
 & 
1389447 & 1325334 & 1263572 & 61796 (04.66\%)  & 11499 (00.87\%) & 
647782 & 34197 & 54401 & 24959 & 26887 \\ 
35...30 & 
1219365 & 1160754 & 1104645 & 56109 (04.83\%)  & 8340 (00.72\%) &
645586 & 32207 & 35775 & 15961 & 18951 \\ 
 & 
1207012 & 1149117 & 1093687 & 67067 (04.82\%)  & 8276 (00.72\%) & 
639412 & 31808 & 35217 & 15961 & 18690 \\ 
30...25 & 
891847 & 849334 & 808569 & 40765 (04.80\%)  & 6528 (00.77\%) &
513495 & 25324 & 23324 & 10344 & 12108 \\ 
 & 
891847 & 849334 & 808569 & 40765 (04.80\%)  & 6528 (00.77\%) & 
513495 & 25324 & 23324 & 10344 & 12108 \\ 
25...20 & 
755711 & 717308 & 680488 & 36820 (05.13\%)  & 4680 (00.65\%) &
483062 & 24705 & 16988 & 6797 & 10043 \\ 
 & 
755711 & 717308 & 680488 & 36820 (05.13\%)  & 4680 (00.65\%) & 
483062 & 24705 & 16988 & 6797 & 10043 \\ 
20...15 & 
546600 & 516452 & 487622 & 28830 (05.58\%)  & 2804 (00.54\%) &
410668 & 20412 & 5113 & 1113 & 4380 \\ 
 & 
546363 & 516230 & 487414 & 29038 (05.58\%)  & 2804 (00.54\%) & 
410479 & 20400 & 5109 & 1113 & 4377 \\ 
15...10 & 
270380 & 252959 & 236444 & 16515 (06.53\%)  & 1206 (00.48\%) &
232277 & 10705 & 1424 & 216 & 1405 \\ 
 & 
270277 & 252859 & 236347 & 16612 (06.53\%)  & 1206 (00.48\%) & 
232182 & 10701 & 1423 & 216 & 1404 \\ 
10...5 & 
102469 & 95058 & 88071 & 6987 (07.35\%)  & 465 (00.49\%) &
90803 & 5806 & 188 & 6 & 187 \\ 
 & 
102469 & 95058 & 88071 & 6987 (07.35\%)  & 465 (00.49\%) & 
90803 & 5806 & 188 & 6 & 187 \\ 
5...0 & 
4742 & 4458 & 4188 & 270 (06.06\%)  & 14 (00.31\%) &
4230 & 443 & 0 & 0 & 0 \\ 
 & 
4742 & 4458 & 4188 & 270 (06.06\%)  & 14 (00.31\%) & 
4230 & 443 & 0 & 0 & 0 \\ 
0...-5 & 
0 & 0 & 0 & 0 (00.00\%)  & 0 (00.00\%) &
0 & 0 & 0 & 0 & 0 \\ 
 & 
0 & 0 & 0 & 0 (00.00\%)  & 0 (00.00\%) & 
0 & 0 & 0 & 0 & 0 \\ 
-5...-10 & 
41107 & 37220 & 33648 & 3572 (09.60\%)  & 201 (00.54\%) &
35622 & 1778 & 70 & 27 & 70 \\ 
 & 
41107 & 37220 & 33648 & 3572 (09.60\%)  & 201 (00.54\%) & 
35622 & 1778 & 70 & 27 & 70 \\ 
-10...-15 & 
279432 & 259483 & 240740 & 18743 (07.22\%)  & 1248 (00.48\%) &
242300 & 9333 & 1122 & 127 & 1099 \\ 
 & 
279425 & 259476 & 240733 & 18750 (07.22\%)  & 1248 (00.48\%) & 
242293 & 9333 & 1122 & 127 & 1099 \\ 
-15...-20 & 
353430 & 330472 & 308636 & 21836 (06.61\%)  & 1711 (00.52\%) &
291263 & 9781 & 1258 & 148 & 1232 \\ 
 & 
353429 & 330471 & 308635 & 21837 (06.61\%)  & 1711 (00.52\%) & 
291262 & 9781 & 1258 & 148 & 1232 \\ 
-20...-25 & 
550777 & 519144 & 488999 & 30145 (05.81\%)  & 3671 (00.71\%) &
414541 & 15475 & 4422 & 905 & 3662 \\ 
 & 
549189 & 517742 & 487767 & 31377 (05.79\%)  & 3667 (00.71\%) & 
413455 & 15391 & 4422 & 905 & 3662 \\ 
-25...-30 & 
838755 & 795233 & 753743 & 41490 (05.22\%)  & 5443 (00.68\%) &
542543 & 23395 & 11374 & 2923 & 7865 \\ 
 & 
837265 & 793810 & 752386 & 42847 (05.22\%)  & 5433 (00.68\%) & 
541721 & 23350 & 11344 & 2923 & 7842 \\ 
-30...-35 & 
846928 & 806816 & 768476 & 38340 (04.75\%)  & 6211 (00.77\%) &
491805 & 23832 & 15389 & 4086 & 11032 \\ 
 & 
840856 & 801185 & 763258 & 43558 (04.73\%)  & 6168 (00.77\%) & 
488416 & 23284 & 15218 & 4086 & 10882 \\ 
-35...-40 & 
740300 & 706978 & 675090 & 31888 (04.51\%)  & 6754 (00.96\%) &
374763 & 19144 & 23143 & 5953 & 17177 \\ 
 & 
740300 & 706978 & 675090 & 31888 (04.51\%)  & 6754 (00.96\%) & 
374763 & 19144 & 23143 & 5953 & 17177 \\ 
-40...-45 & 
593885 & 568458 & 544091 & 24367 (04.29\%)  & 5218 (00.92\%) &
264591 & 15619 & 21078 & 6762 & 12516 \\ 
 & 
593885 & 568458 & 544091 & 24367 (04.29\%)  & 5218 (00.92\%) & 
264591 & 15619 & 21078 & 6762 & 12516 \\ 
-45...-50 & 
531705 & 506589 & 482601 & 23988 (04.74\%)  & 5092 (01.01\%) &
213189 & 13257 & 19894 & 7570 & 9858 \\ 
 & 
531705 & 506589 & 482601 & 23988 (04.74\%)  & 5092 (01.01\%) & 
213189 & 13257 & 19894 & 7570 & 9858 \\ 
-50...-55 & 
490341 & 466454 & 443617 & 22837 (04.90\%)  & 5209 (01.12\%) &
188403 & 12471 & 15447 & 5210 & 9582 \\ 
 & 
490341 & 466454 & 443617 & 22837 (04.90\%)  & 5209 (01.12\%) & 
188403 & 12471 & 15447 & 5210 & 9582 \\ 
-55...-60 & 
555550 & 529816 & 505064 & 24752 (04.67\%)  & 7388 (01.39\%) &
175366 & 14785 & 25487 & 10287 & 14607 \\ 
 & 
555550 & 529816 & 505064 & 24752 (04.67\%)  & 7388 (01.39\%) & 
175366 & 14785 & 25487 & 10287 & 14607 \\ 
-60...-65 & 
495619 & 471024 & 447362 & 23662 (05.02\%)  & 4956 (01.05\%) &
138722 & 12780 & 31257 & 14079 & 15920 \\ 
 & 
495619 & 471024 & 447362 & 23662 (05.02\%)  & 4956 (01.05\%) & 
138722 & 12780 & 31257 & 14079 & 15920 \\ 
-65...-70 & 
339441 & 324640 & 310346 & 14294 (04.40\%)  & 3697 (01.14\%) &
97779 & 8445 & 9454 & 3211 & 6149 \\ 
 & 
339441 & 324640 & 310346 & 14294 (04.40\%)  & 3697 (01.14\%) & 
97779 & 8445 & 9454 & 3211 & 6149 \\ 
-70...-75 & 
176928 & 170122 & 163529 & 6593 (03.88\%)  & 2036 (01.20\%) &
47308 & 4299 & 5095 & 2569 & 2264 \\ 
 & 
176928 & 170122 & 163529 & 6593 (03.88\%)  & 2036 (01.20\%) & 
47308 & 4299 & 5095 & 2569 & 2264 \\ 
-75...-80 & 
66080 & 63965 & 61904 & 2061 (03.22\%)  & 550 (00.86\%) &
17753 & 1794 & 194 & 10 & 194 \\ 
 & 
66080 & 63965 & 61904 & 2061 (03.22\%)  & 550 (00.86\%) & 
17753 & 1794 & 194 & 10 & 194 \\ 
-80...-85 & 
35054 & 34072 & 33110 & 962 (02.82\%)  & 315 (00.92\%) &
9097 & 884 & 192 & 9 & 192 \\ 
 & 
35054 & 34072 & 33110 & 962 (02.82\%)  & 315 (00.92\%) & 
9097 & 884 & 192 & 9 & 192 \\ 
-85...-90 & 
3564 & 3461 & 3360 & 101 (02.92\%)  & 30 (00.87\%) &
1118 & 108 & 0 & 0 & 0 \\ 
 & 
3564 & 3461 & 3360 & 101 (02.92\%)  & 30 (00.87\%) & 
1118 & 108 & 0 & 0 & 0 \\ 
90...-90 & 
23310532 & 22207563 & 21147514 & 1060049 (04.77\%)  & 192727 (00.87\%) &
10167460 & 577544 & 860224 & 388222 & 426992 \\ 
 & 
23267132 & 22166709 & 21109088 & 1098475 (04.77\%)  & 192448 (00.87\%) & 
10149170 & 575968 & 858316 & 388222 & 426154 \\ 
\enddata
\tablenotetext{a}{For each Galactic latitude range, the second row gives counts excluding sources in extended objects such as galaxies or stellar clusters larger than 30arcmin (see \citet{bia17guvcat}). 
Counts for sources with significant detection also in GALEX  $FUV$ are indicated with "FUV". "Point-like" sources are defined by the SDSS morphological classification of $TYPE$="STAR", that includes both galactic and extra-galactic point-like sources.
    }
\end{deluxetable}

\end{document}